\documentclass[10pt,a4paper]{article}

\usepackage[utf8]{inputenc}
\usepackage[T1]{fontenc}

\usepackage{bbm}
\usepackage{hyperref}
\usepackage{amsmath}
\usepackage{amsfonts}
\usepackage{amsthm}
\usepackage{amssymb}
\usepackage{graphicx}

\newcommand{\N}{\mathbb N}

\begin{document}

\title{Damping effect in innovation processes: case studies from Twitter}

\author{ Giacomo Aletti
\footnote{ADAMSS Center,
  Universit\`a degli Studi di Milano, Milan, Italy, giacomo.aletti@unimi.it}
$\quad$and$\quad$ 
Irene Crimaldi
\footnote{IMT School for Advanced Studies, Lucca, Italy, 
irene.crimaldi@imtlucca.it}
} 
\maketitle

\abstract{
Understanding the {\em innovation} process, that is the underlying mechanisms through which novelties emerge, diffuse and trigger further novelties is undoubtedly of fundamental importance in many areas (biology, linguistics, social science and others). The models introduced so far satisfy the Heaps' law, regarding the rate at which novelties appear, and the Zipf's law, that states a power law behavior for the frequency distribution of the elements. However, there are empirical cases far from showing a pure power law behavior and such a deviation is present for elements with high frequencies.  We explain this phenomenon by means of a suitable {\em ``damping'' effect} in the probability of a repetition of an old element. While the proposed model is extremely general and may be also employed in other contexts, it has been tested on some Twitter data sets and demonstrated great performances with respect to Heaps' law and, above all, with respect to the fitting of the frequency-rank plots for low and high frequencies.}\\[10pt]

%
%
Heaps' law, Innovation process,
Poisson-Dirichlet process, P\'olya urn, preferential attachment,
species sampling sequence, Twitter, Zipf's law

\section*{Introduction: framework and scope}

In our lives we continuously perform actions and these actions can be
the repetition of something we have already done in the past or they
can be a new experience: we can employ a technology that we already
know or we can decide to try a new one, we can listen again a song
that we already listened to in the past or we can decide to listen a
new song, we can see old friends or we can decide to meet new people
and so on. As a consequence, with our actions, we can contribute to
diffuse an existing word or idea or product, or we can create a new
trend. In particular, thinking about social platform, like {\it
  Twitter}, users can diffuse an existing post by means of a
``retweet'' or a ``quote'' of it, or they can write a new one.\\

\indent Understanding the {\em innovation} process, that is the
underlying mechanisms through which novelties emerge, diffuse and
trigger further novelties is undoubtedly of fundamental importance in
many areas (biology, linguistics, social science and others
\cite{armano, arthur, fink, gooday, obrien, puglisi, reader, rogers,
  rzh, saracco, sole, thurner}). Novelties can be viewed as first time
occurrences of some event and the mathematical object used to model an
innovation process is an {\em urn model with infinitely many colors},
also known as {\em species sampling sequence} \cite{han-pitman,
  pitman_1996, Zabell_1992}. Let $X_1$ the first observed color, then,
given the colors $X_1,\dots, X_t$ of the first $t$ extractions, the
color of the $(t+1)$-th extracted ball is a new one with a probability
$b_t$ which is a function of $X_1,\dots,X_t$ (sometimes called ``birth
probability'') and it is equal to the already observed color $c$ with
probability $p_{t,c}=\sum_{n=1}^t q_{t,n}I_{\{X_n=c\}}$, where
$q_{t,n}$ is a function of $X_1,\dots,X_t$. The quantities $b_t$ and
$q_{t,n}$ specify the model: precisely, $b_t$ describes the
probability of having a new color (that is a novelty) at time-step
$t+1$ and $q_{t,n}$ is the weight at time-step $t$ associated to the
event $n$, with $1\leq n\leq t$, so that the probability of having at
time-step $t+1$ the ``old'' color $c$ is proportional to the total
weight at time-step $t$ associated to that color (a reinforcement
mechanism, called ``weighted preferential attachment'' principle).
Note that the number of possible colors is not fixed a priori, but new
colors continuously enter the system. We can see the urn with
infinitely many colors as the space of possibilities, while the
sequence of extracted balls with their colors represents the history
which has been actually realized.  \\

\indent Although there are only a few explicit prediction rules which
give rise to exchangeable sequences, this kind of prediction rules are
widely used, because {\em exchangeability} is a natural assumption in
many statistical problems, in particular from the Bayesian viewpoint,
and many theoretical results are known for exchangeable sequences
\cite{ pitman_1995, pitman, Zabell_1992}. We recall that a sequence is
said exchangeable if its joint distribution is invariant with respect
to permutations that act on only finitely many indices, with the rest
fixed. Exchangeability is a powerful assumption, but, in some
situations, it could be too restrictive and unrealistic, because it
does not take into account the possible causality in the
data. Therefore the introduction and study of species sampling
sequences, which are not exchangeable, but which still have
interesting theoretical properties, is welcome.  \\

\indent The {\em Blackwell-MacQueen urn} scheme \cite{black-mac,
  pitman_1996} provides the most famous example of exchangeable
prediction rule.  According to this prediction rule, a new color is
observed with probability $b_t=\theta/(\theta+t)$, where $\theta>0$,
and an old color is observed with a probability proportional to the
number $K_{c,t}$ of times that color was extracted in the previous
extractions: $q_{t,n}=1/(\theta+t)$,
i.e.~$p_{c,t}=K_{c,t}/(\theta+t)$.  This is the ``simple''
preferential attachment rule, also called ``popularity''
principle. This urn model is also known as Dirichlet process
\cite{ferguson} or as Hoppe's model~\cite{hoppe} and, in terms of
random partitions, it corresponds to the so-called {\em Chinese
  restaurant process} \cite{pitman}. Afterwards, it has been extended
introducing an additional parameter and it has been called {\em
  Poisson-Dirichlet model}~\cite{james, pitman, pit-yor,
  Teh2006}. More precisely, for the Poisson-Dirichlet model, we have
\begin{equation*}
  \begin{split}
  &b_t=\frac{\theta+\gamma D_t}{\theta +  t},
  \qquad q_{t,n}=\frac{1-\gamma/K_{X_n,t}}{\theta + t},
  \\
&\mbox{and so }\quad
p_{c,t}=
\frac{K_{c,t}-\gamma}{\theta +  t},
\end{split}
\end{equation*}
where $0\leq \gamma<1$, $\theta>-\gamma$ and $D_t$ denotes the number
of distinct extracted colors until time-step $t$. This model again
generates an exchangeable sequence. In \cite{BaCrLe}, the authors
introduce and study a generalization of the Poisson-Dirichlet urn,
introducing some random weights so that an old color is observed with
a probability proportional to the total weight associated to that
color during the previous extractions. The model so obtained does not
give rise to an exchangeable sequence anymore, but the generated
sequence is a conditionally identically distributed sequence
\cite{ber-pra-rig} and so properties usually required in Bayesian
statistics are preserved.  \\

\indent From an applicative point of view, as an innovation process,
the Poisson-Dirichlet process has the merit to reproduce in many cases
the correct basic statistics, namely the Heaps'~\cite{heaps_1978,
  herdan_1960} and the (generalized) Zipf's
laws~\cite{zipf_1929,zipf_1935,zipf_1949}, which quantify,
respectively, the~rate at which new elements appear and the frequency
distribution of the elements.  \\

\indent The {\em Heaps' law} states that the number $D_t$ of distinct
observed elements (i.e.~colors, according the metaphor of the urn)
when the system consists of $t$ elements (i.e.~after $t$ extractions
from the urn) follows a power law: $D_t\propto t^\gamma$, $0<\gamma
\leq 1$.  Recently, Tria et al.~\cite{Tria3, Tria1, Tria2} have
introduced and studied a new model, called {\em urn with triggering},
that includes the Poisson-Dirichlet process as a particular case. This
model is based on Kauffman's principle of the adjacent
possible~\cite{kauffman_2000}: indeed, the model starts with an urn
with a finite number $N_0>0$ of balls with distinct colors and,
whenever a color is extracted for the first time, a set of balls with
new colors is added to the urn. This represents Kauffman's idea that,
when a novelty occurs, it triggers further novelties. Therefore, in
the urn with triggering, the space of possible colors expands and it
can be seen as an urn with infinitely many colors, where
\begin{equation*}
b_t=\frac{N_0+\nu D_t}{N_0+\rho t+ a D_t},
\qquad\mbox{and}\qquad 
p_{c,t}=\frac{\rho K_{c,t}+ a-\nu}{N_0+\rho t+ a D_t},
\end{equation*}
where $\nu \geq 0$, $\rho > 0$ and $a=\widetilde\rho+\nu-\rho+1$ with
$\widetilde\rho> -1$. The Poisson-Dirichlet model corresponds to the
case $a=0$ (taking $\theta=N_0/\rho$ and $\gamma=\nu/\rho\in
[0,1)$). In general, the sequences generated by the urn with
  triggering are not exchangeable. Moreover, while the
  Poisson-Dirichlet process can predict only a sub-linear power law
  behavior for the number of distinct observed colors/elements
  (i.e. Heaps' law with~$\gamma<1$), the urn with triggering is able
  to provide also a linear growth for it (i.e. Heaps' law
  with~$\gamma=1$): precisely, we have $D_t\propto t^{\nu/\rho}$ for
  $0<\nu<\rho$, $D_t\propto t/\ln(t)$ for $\nu=\rho$, $D_t\propto t$
  for $\nu>\rho$ and, finally, $D_t\propto\ln(t)$ for $\nu=0$. A
  different model able to reproduce all the Heaps' exponents in
  $[0,1]$ is the one proposed in \cite{Latora2018}, where the
  innovation process is described as an edge-reinforced random walk on
  an underlying network of relations among concepts (nodes). The
  topology naturally plays a key role in this last model so that it
  results useful only in the cases where the network can be well
  reconstructed from data (e.g. innovations in a scientific discipline
  by means of the analysis of scientific publications).  \\

\indent The {\em Zipf's law} states an inverse proportionality between
the frequency and rank of the considered quantities.  Let us consider
a generic sequence of elements (colors) and count the number of
occurrences of each element. Now, suppose one repeats the same
operation for all the distinct elements in the sequence, and ranks all
the elements according to their frequency of occurrence (rank $r=1$
corresponds to the most frequent color, the rank $r=2$ correspond to
the second most frequent color and so on, the higher the rank, the
less frequent the color) and plots them in a graph showing the number
of occurrences versus the rank. The (generalized) Zipf's law affirms a
frequency-rank distribution of the form $z(r)\propto r^{-\alpha}$,
with~$0<\alpha<+\infty$ (the strict Zipf's law refers to $\alpha
=1$). This property is used in theoretical analyses \cite{lu_2010} or
it is seeked by mean of various mechanisms
\cite{Corominas-Murtra_2015, Cubero:2018zyp, mitzen, newman_2005,
  simon_1955, zanette2005}. The urn with triggering asymptotically
(i.e. for large times) satisfies the Zipf's law when $\nu\neq 0$, with
the relationship between the Heaps' and Zipf's exponents given by
$\gamma = 1/\alpha$ when $\gamma=\nu/\rho<1$. \\

\indent However, in some cases the frequency-rank plots observed in
empirical applications are far from showing a pure power law behavior
and the above relation between the Heaps' and Zipf's exponents holds
asymptotically and only when looking at the tail of the frequency-rank
plot, i.e., for large ranks, that correspond to small frequencies
(rare elements) \cite{cancho, cattuto, ger, lu2013, petersen, Tria1,
  Tria2}. To the best of our knowledge, few works have tried to
explain the empirical frequency-rank plots in the part of small ranks,
i.e. for high frequencies, where they typically deviate from a power
law behavior. The papers \cite{cancho, ger, petersen} provides
examples where the frequency-rank plot exhibits a power law behavior
with two different scaling exponents: $z(r)\propto r^{-\alpha_1}$ for
$r< \xi$ and $z(r)\propto r^{-\alpha_2}$ for $r>\xi$, where $\alpha_1$
is typically equal to $1$.  In \cite{ger} this empirical finding is
achieved by means of a generative stochastic model based on the
existence of two different classes of words: a finite number of core
words, which are more frequently used and do not affect the
probability of a new word to appear and the remaining infinite number
of non-core words, which are significantly less used and, once used,
reduce the probability of a new word to be employed in the future. In
\cite{petersen} no generative models are provided, but the two scaling
regimes of the word frequency distributions are again explained using
two categories of words: the specialized words in the ``unlimited
lexicon'' are not universally shared and are employed less frequently
than the words in the ``kernel lexicon''. In \cite{cattuto}, the
authors introduce a variant of the Simon's model \cite{simon_1955}
(which is a species sampling sequence where $b_t$ is equal to a
constant and the probability of observing an old color is proportional
to the number of times that color was extracted in the previous
extractions) that also incorporates a long-term memory or aging
component: precisely, at a time step $t$, a new element appears with
probability $b$, whereas with probability $(1-b)$, an old element is
chosen, going back in time by $n$ time steps with a probability
$Q_t(n)$ that decays as a power law. This generative stochastic model
is able to fit with high accuracy the observed frequency-rank plots in
the collaborative tagging context.  Finally, in \cite{lu2013}, the
authors observe an exponential decay in the frequency-rank plot and
this fact is ascribed to the limited dictionary size (indeed, they
consider Chinese, Japanese and Korean languages). In order to
reproduce this empirical finding, they provide a generative model
based on a finite dictionary size of distinct characters.  \\

\indent In this work we are going to show that the deviations from the
Zipf's law in the empirical frequency-rank plots (in particular, in
the part of small ranks) can be explained adding a {\em ``damping''
  effect} in the urn with triggering.  More precisely, we generalize
the urn with triggering model by the introduction of a function $F$
that drives the update mechanism of the number of balls of the same
color of the extracted one when it is of an ``old'' color. In the
standard model, this function is linear so that it generates a power
law behavior of the frequency-rank plot (Zipf's law), that usually
matches the empirical ones only in the part of rare elements
(i.e.~large ranks). Instead, if we take the function $F$ linear until
a certain point and then still linear but with a smaller slope or
sub-linear (for instance, the square root), then we obtain a
frequency-rank plot closer to the empirical ones also in the part of
high frequencies. This fact can be seen as a damping effect on the old
elements: the number of balls of an old color increases linearly with
the number of times it is extracted until a certain threshold, then it
increases slower. Our unique general model is able to reproduce the
empirically observed power law behavior with two different scaling
exponents mentioned above and also other kinds of curves, observed in
real data sets. Indeed, given the function $g$ that fits the empirical
frequency-rank plot (see eq.~\eqref{eq:Zipf}), we are able to find the
corresponding function $F$ of the proposed generative model (see
eq.\eqref{rel-F-g}). This is a very useful result for applicative
purposes, since in applications one usually observes and tries to fit
the empirical frequency-rank plot. Further, we have shown how to
obtain the asymptotic behavior of the number $D_t$ of distinct
observed elements starting from the function $g$ and we have employed
this methodology with some specific functions $g$. The obtained
theoretical results are supported by simulations.\\

\indent We apply the proposed model to some data sets from the social
platform {\it Twitter}. The main mechanism in Twitter leading content
diffusion is the possibility for users to ``re-tweet'', reply or quote
the post sent by others. Therefore, ordered sequences of posts can be
seen as generated by an urn model with infinitely many colors: a new
color is associated to a new tweet, while the extraction of an old
color corresponds to a re-sharing (by means of a re-tweet or a quote
or a reply to) of an old post. The update function $F$ rules the
probability of a generic posted tweet to be re-shared. For all the
considered data sets, we observe the same damping effect on the old
elements: the update function $F$ grows linearly until a certain
threshold, then it increases sub-linearly, precisely according to the
square root. Moreover, we empirically verify the linear growth of the
variable $D_t$, which agrees with the proven theoretical result. We
refer to \cite{guille-survey} for a survey on information diffusion in
on-line social networks. In \cite{lambiotte} the authors face the
problem to predict the future time evolution of the popularity of a
certain tweet and, in particular, to estimate the final number of
retweets of that given tweet (see also the reference therein for other
works related to the same question). In \cite{thij-2016, thij-2014}
the focus is instead on the ``retweet graph'', which is the graph of
users who participated in the discussion of a specific topic and where
a directed edge indicates that a user retweeted a tweet of another
user.\\

\indent We underline that the proposed model is very general and
flexible and so it may be also employed in other contexts. Finally, we
explain our choice of the term ``damping'' with respect to the other
terms ``saturation'' and ``aging'' employed in literature. The term
``saturation'' is typically used when the probability of observing a
new element goes to zero in such a way that $D_t$ converges to a
certain finite value \cite{lu2013}; while, the term ``aging'' refers
to time in the sense that the probability of having the repetition of
a certain old element decreases with the difference between the
present time and the last time of observation of that old element
\cite{cattuto}. The effect that we model refers to a damping factor in
the probability of a repetition of an old element: using the metaphor
of the urn, the number of balls of a given color in the urn increases
with the number of times that color has been extracted according to a
suitable update function that exhibits two different speeds, one
before a certain threshold and a lower one after the
threshold. Therefore, it obviously differs from the above saturation
effect and it is related in some sense to the age of the elements, but
not directly and so it is not exactly the same of the above recalled
aging effect.

\section{Results}\label{sec-results}
We firstly introduce the model, using the metaphor of the urn, and we
state the main related results. Then we show the empirical results.
\\

\noindent{\bf Model.} Given an increasing function $F$ defined on
${\mathbb\N}\setminus\{0\}$ with $F(1)>0$, the model works as follows.
An urn initially contains balls of $N_0>0$ different colors. For each
color, we have one ball. Then, at~each time step $t$, a~ball is drawn
at random from the urn, its color is registered in a sequence
$\mathcal{S}$ and:
\begin{itemize}
\item if the color of the extracted ball is a new one, that is it
  appears for the first time in $\mathcal{S}$ (it corresponds to the
  realization of a novelty), then the number of balls of the extracted
  color in the urn becomes $F(1)>0$ and we add $\nu+1$ (with $\nu\geq
  0$) distinct balls of different new colors, that is of colors not
  yet present in the urn: precisely, we add one ball for each new color;
\item if the color $c$ of the extracted ball is already present in
  $\mathcal{S}$ and $K_{c,t}$ denotes the number of times the color
  $c$ was extracted until time step $t$ (included), the number of
  balls of color $c$ in the urn becomes $F(K_{c,t})>0$.
\end{itemize}
The fact that the {\em update function} $F$ is increasing means that
we have a {\em reinforcement mechanism}: the larger the number of
times the color $c$ has been extracted, the larger the number of balls
of color $c$ in the urn. Moreover, the addition of a set of balls with
new colors in the urn whenever a color is extracted for the first time
represents Kauffman's {\em principle of the adjacent
  possible}~\cite{kauffman_2000}, i.e.~ the idea that, when a novelty
occurs, it triggers further novelties.  \\

\indent If~$X_{t+1}$ is the color of the extracted ball at time step
$t+1$, $D_t$ is the number of different colors extracted until time
step $t$, $N_t$ is the number of different colors in the urn until
time step $t$ and $T_t$ is the total number of balls in the urn at
time step $t$, we have:
 \begin{equation*}
   \begin{split}
N_t&=N_0+(\nu+1)D_t\,\\
T_t&=\sum_{j=1}^{N_t} F(K_{j,t})=
N_0+\nu D_t+\sum_{j\in \mathcal{S}} F(K_{j,t})
  \end{split}
 \end{equation*}
(the sum $\sum_{j\in \mathcal{S}}$ denotes the sum on the $D_t$
 different colors in $\mathcal S$) and
\begin{equation}\label{eq:b_t}
b_t=P(X_{t+1}=\hbox{\em{new}}\,|\,X_1,\dots,X_t)=
\frac{(N_t-D_t)}{T_t}
=
\begin{cases}
  1\qquad&\mbox{for } t=0\\
  \frac{N_0+\nu D_t}{N_0+\nu D_t+\sum_{j\in \mathcal{S}} F(K_{j,t})}
  \quad&\mbox{for } t\geq 1\,.
  \end{cases}
\end{equation}
Moreover, if~$c$ denotes an {\em old} color, we have for each $t\geq 1$
\begin{equation}\label{eq:p_ct}
  p_{c,t}=P(X_{t+1}=c\,|\,X_1,\dots,X_t)=
  \frac{F(K_{c,t})}{T_t}=
\frac{F(K_{c,t})}{N_0+\nu D_t+\sum_{j\in \mathcal{S}} F(K_{j,t})}\,.
\end{equation}
The quantities $b_t$ and $p_{c,t}$ drive the model. The model
parameters $N_0$ and $\nu$ can be any real positive numbers with
$\max\{N_0,\nu\}>0$. The update function $F$ can be any increasing
function defined on the strictly positive integer numbers and with
values in $(0,+\infty)$. For instance, the standard urn model with
triggering \cite{Tria3, Tria1, Tria2} corresponds to the update
function $F$ defined as $F(x)= \widehat{\rho}+\rho(x-1)$, with
$\widehat{\rho}=\widetilde{\rho}+1>0$ and $\rho>0$.
%
%
\\

\indent For the case $\nu>0$, when we assume a dependence in the
frequency-rank plot of the form $g(z(r)) = -a \ln(r) + b$, with an
invertible differentiable function $g$ and a constant $a>0$, we obtain
(see the appendix for the computations) the relation
\begin{equation}\label{rel-F-g}
F(x)\approx \frac{a\nu}{g'(x)}.
\end{equation}
This relation is not an exact equality, because it is due to
\eqref{eq:master2} and \eqref{eq:master3}, which have been obtained by
approximation. From the applicative point of view, relation
\eqref{rel-F-g} is of fundamental utility because it allows us to
guess the right update function $F$ in the model by means of the
function $g$ that fits well the empirical frequency-rank plot. For
instance, for the standard urn with triggering, we have $g=\ln$ with
$a=\rho/\nu$ and, indeed, we have $F(x)=\widehat\rho+\rho(x-1)\approx
\rho x=a\nu x$ for large $x$ and so \eqref{rel-F-g} is satisfied. More
generally, whenever we have a Zipf's law, that is $g=\ln$ in
\eqref{eq:Zipf}, we get that, according to \eqref{rel-F-g}, we have to
choose a linear update function $F$ in the model.  Regarding the
asymptotic behavior of the number $D_t$ of distinct elements in
$\mathcal S$ as a function of its length $t$, we note that
\begin{equation}\label{rel-Dt}
t=\sum_{r=1}^{D_t} z(r)\approx \int_1^{D_t}z(r)dr
\end{equation}
and so the behavior of $D_t$ can be obtained starting from the
frequency-rank function $r\mapsto z(r)$. In the appendix, we derive
the asymptotic behavior of $D_t$ starting from different kinds of
function $g$. \\

\begin{figure}[htbp]
\begin{center}
\includegraphics[width=0.47\textwidth]{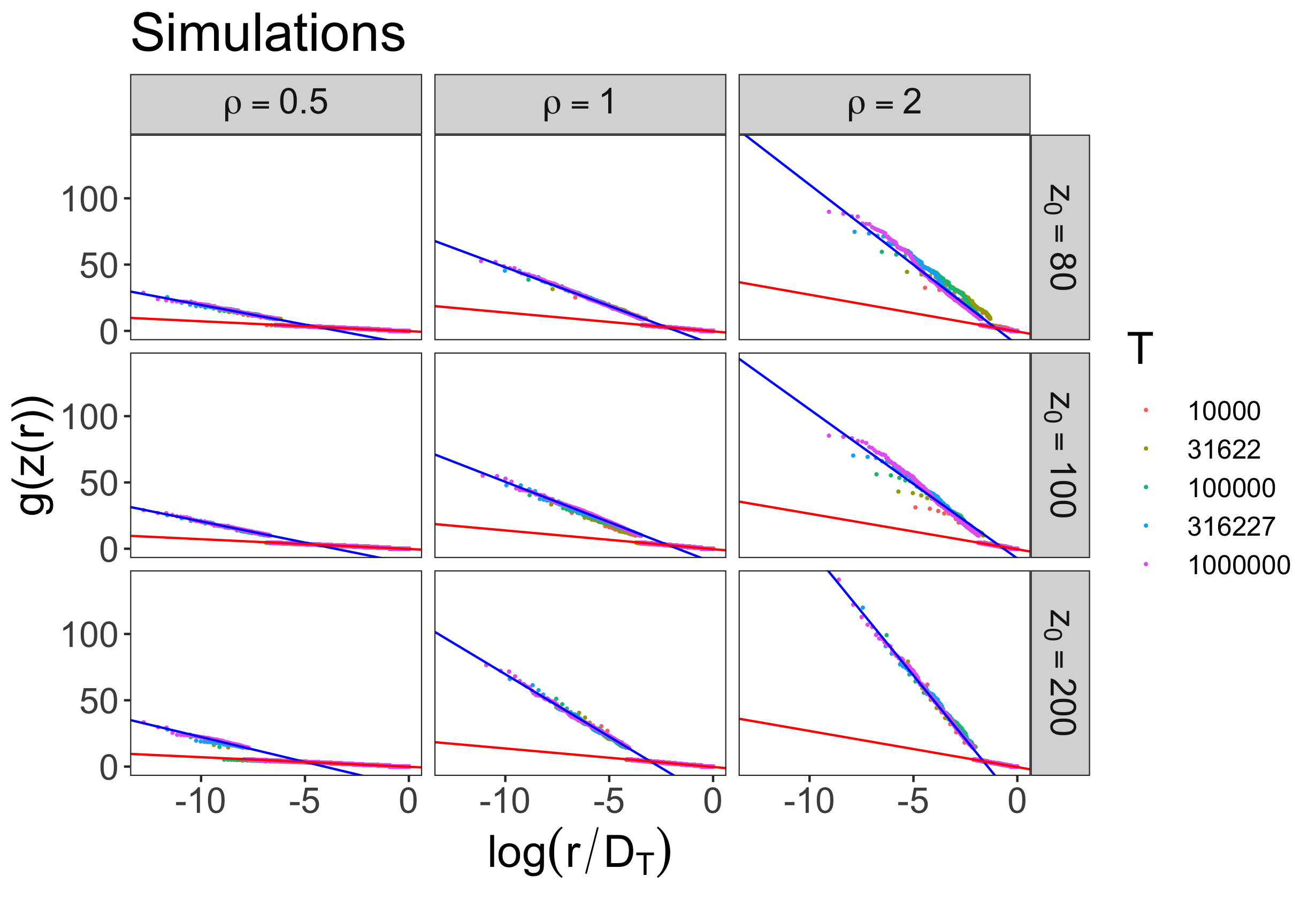}
 \includegraphics[width=0.47\textwidth]{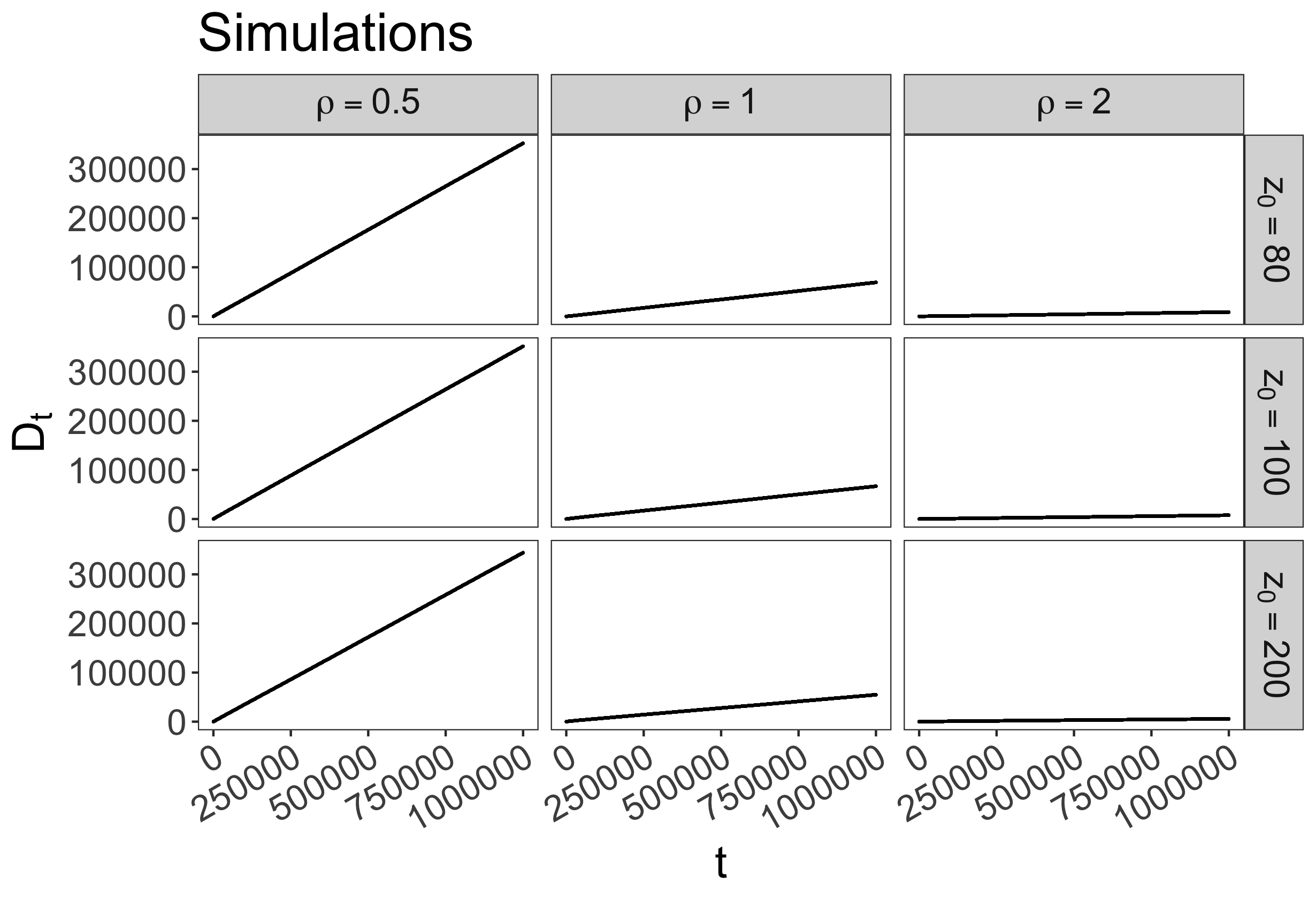}
\end{center}
	\caption{Simulations for case \eqref{F-sqrt} with $N_0=2$,
          $\nu=0.75$ and different values of $\rho$ and $z_*=z_0$:
          (Left) Frequency-rank plot in two different scales: log-sqrt
          before a certain $R_*$ and log-log after $R_*$. The
          different colors of the dots correspond to different
          quantities of data taken for depicting the plot. (Right)
          Behavior of $D_t$.}
\label{sim-sqrt}
\end{figure}

\begin{figure}[htbp]
\begin{center}
\includegraphics[width=0.47\textwidth]{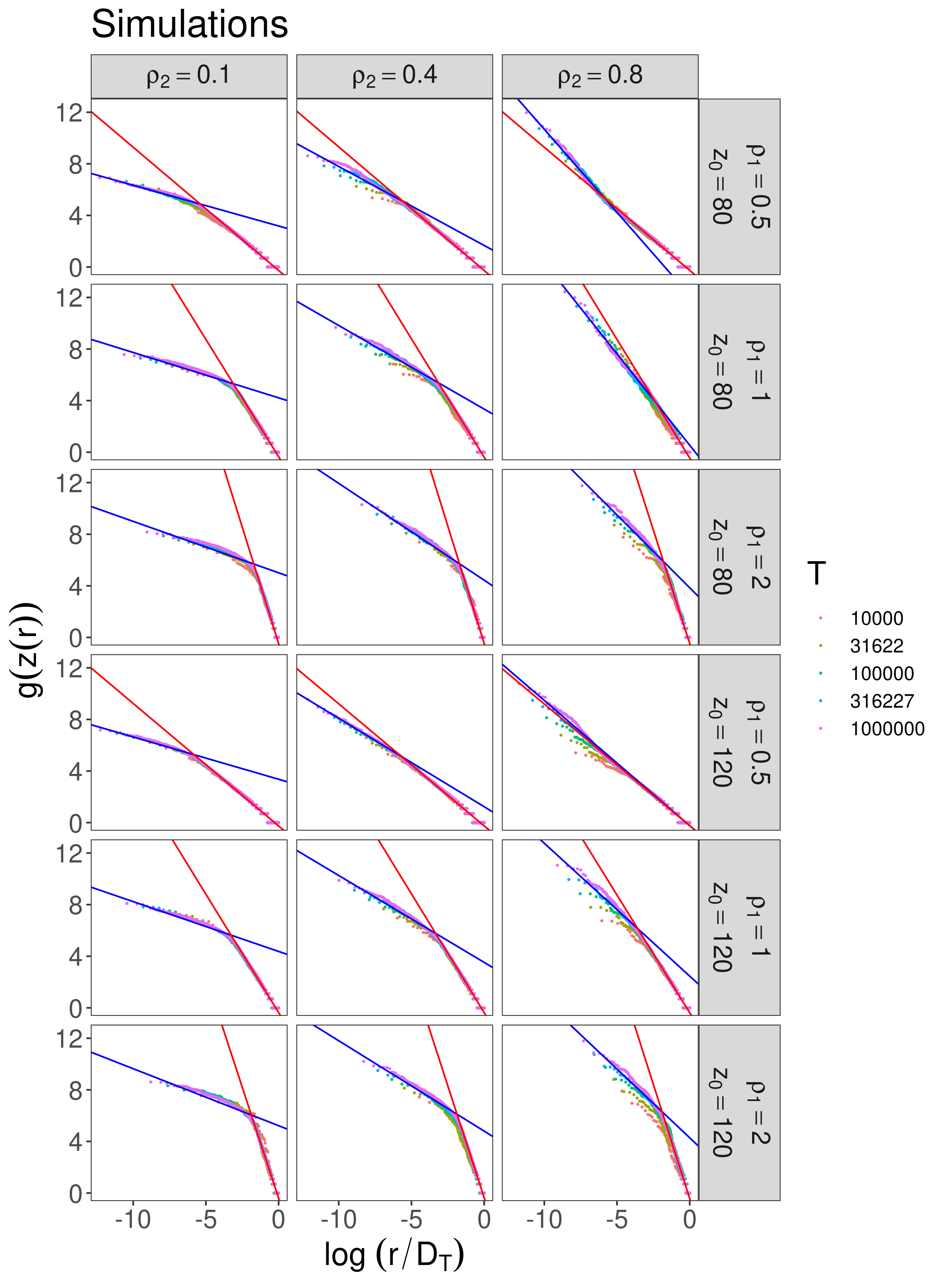}
  \\
\includegraphics[width=0.47\textwidth]{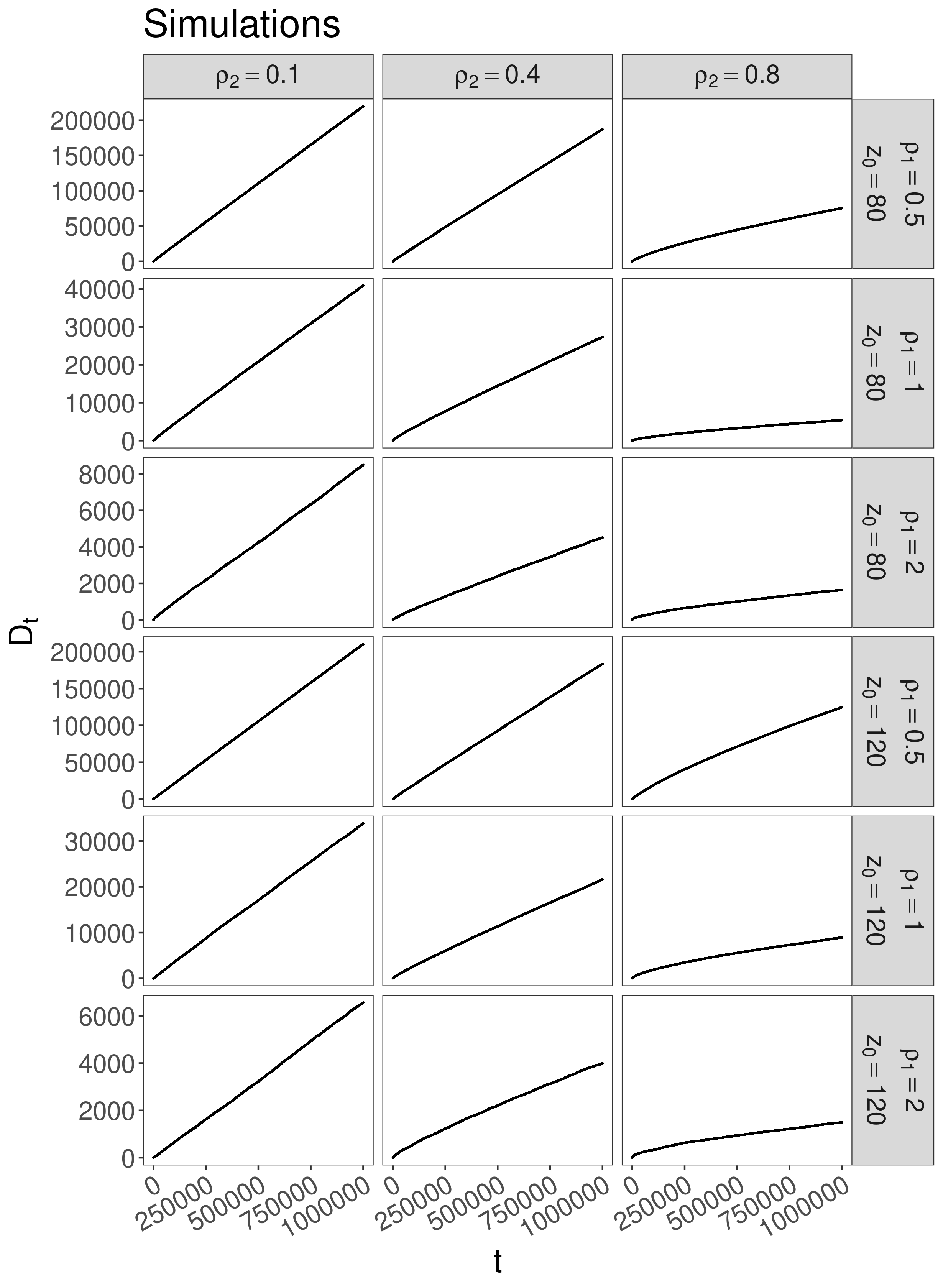}
\includegraphics[width=0.47\textwidth]{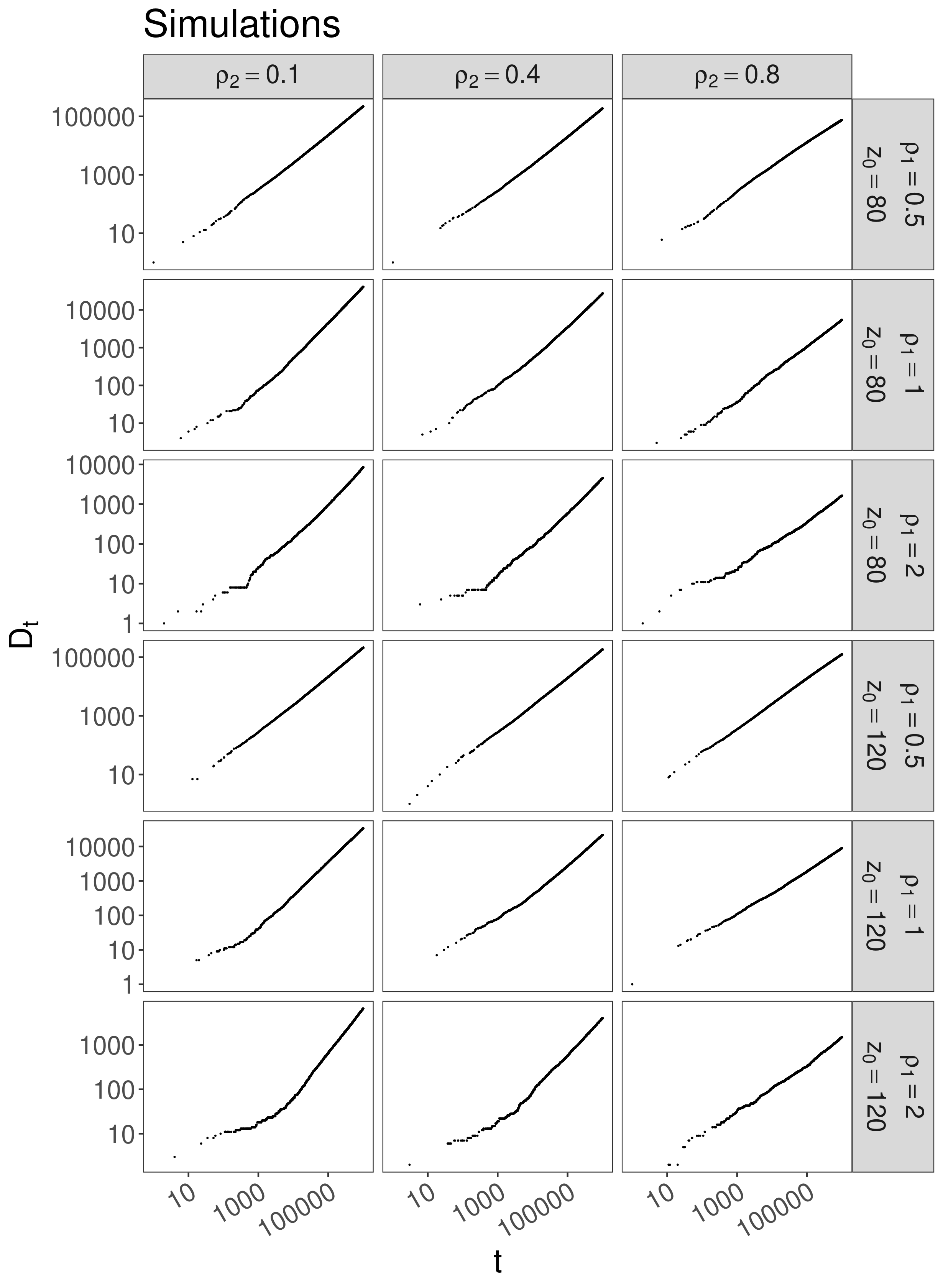}
\end{center}
	\caption{Simulations for case \eqref{F-double-lin} with
          $N_0=2$, $\nu=0.75$ and different values of
          $\rho_1,\,\rho_2$ and $z_*=z_0$: (Up) Frequency-rank plot in
          log-log scale. The different colors of the dots correspond
          to different quantities of data taken for depicting the
          plot. (Below) Behavior of $D_t$: the plot on the left shows
          a linear growth of $D_t$ (Heaps' law with exponent $1$) when
          $\rho_2/\nu<1$ and the plot on the right shows a power-law
          growth (Heaps' law with exponent smaller than $1$) when
          $\rho_2/\nu>1$.}
\label{sim-double-lin}
\end{figure}

\begin{figure}[htbp]
\begin{center}
\includegraphics[width=0.47\textwidth]{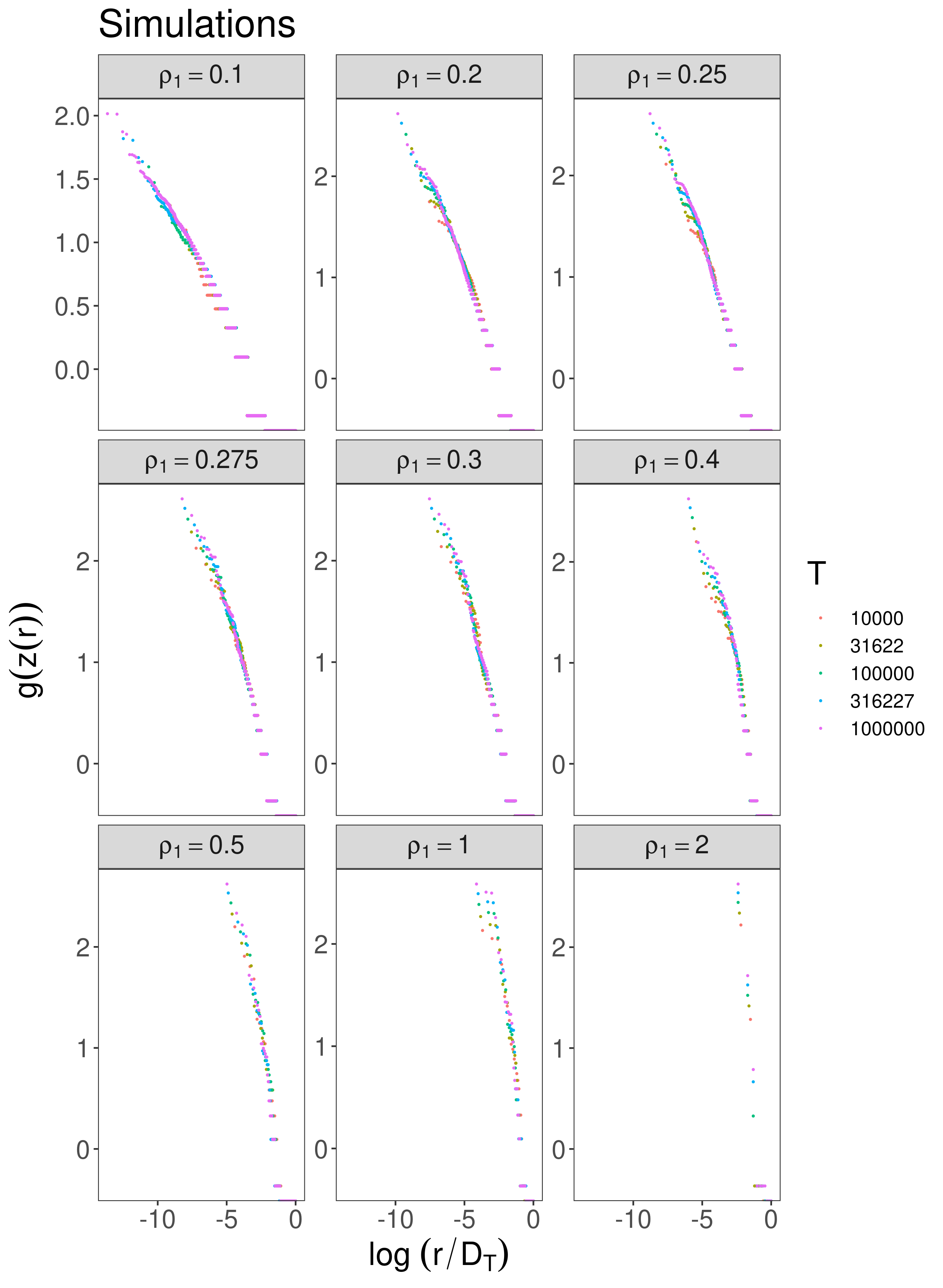}
 \includegraphics[width=0.47\textwidth]{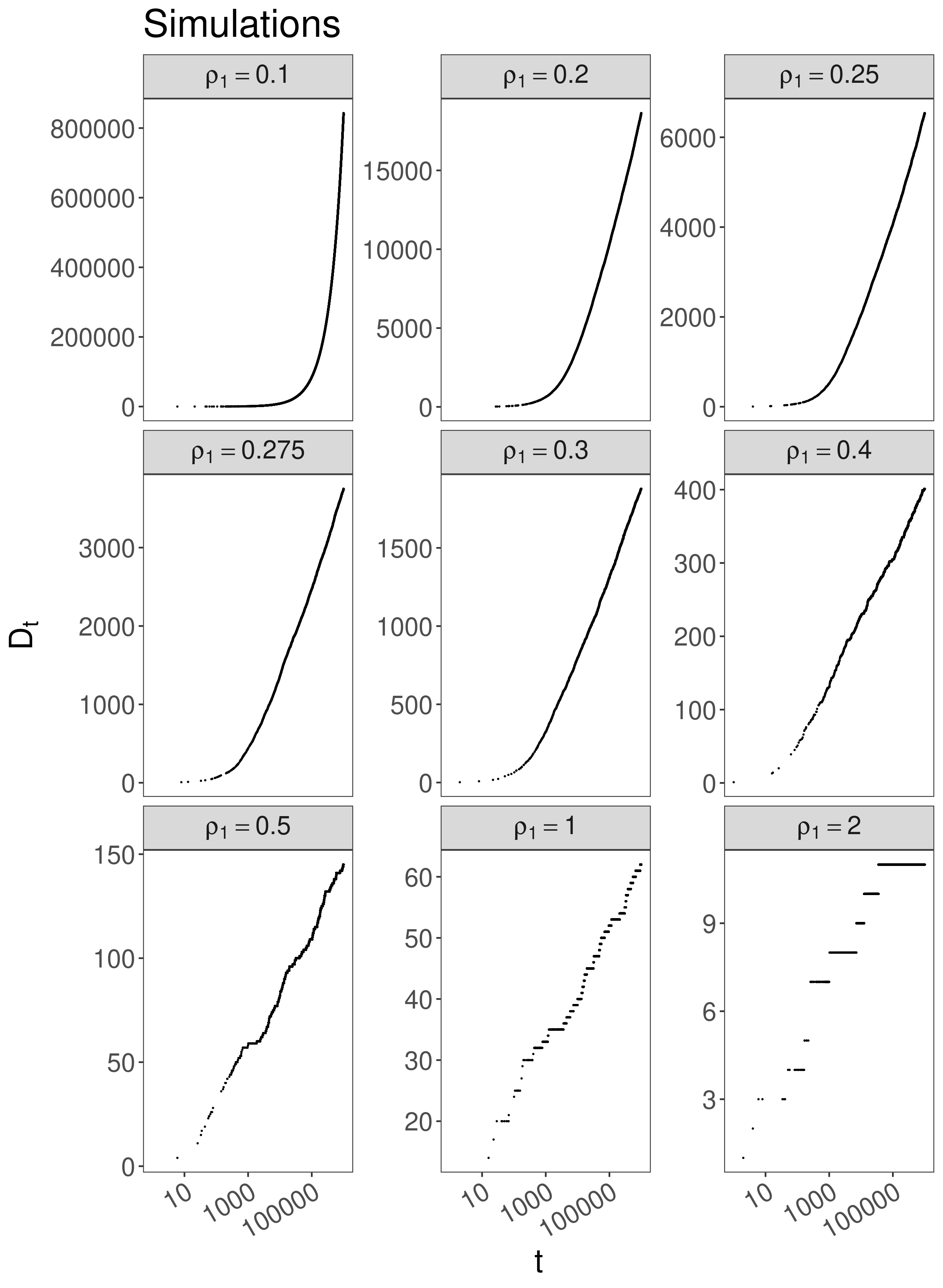}
\end{center}
	\caption{Simulations for case \eqref{F-xlog} with $N_0=2$,
          $\nu=0.75$ and different values of $\rho_1$: (Left)
          Frequency-rank plot with $g(z)=\ln(\ln(z))$. The different
          colors of the dots correspond to different quantities of
          data taken for depicting the plot. (Right) Behavior of
          $D_t$.}
\label{sim-xlog}
\end{figure}

\indent In Figures \ref{sim-sqrt}, \ref{sim-double-lin} and
\ref{sim-xlog}, we show the simulations of the model with $N_0=2$,
$\nu=0.75$ and different kinds of update function $F$. The relation
\eqref{rel-F-g} and the other theoretical results described in the
appendix are supported by these simulations. More precisely, in
Fig.~\ref{sim-sqrt} we exhibit the results for
\begin{equation}\label{F-sqrt}
F(x)=\begin{cases}
\rho  x  \quad&\mbox{for } 1\leq x<z_*\\
\rho \sqrt{z_*x}  \quad&\mbox{for } x\geq z_*\,,
\end{cases}
\end{equation}
for different values of $\rho$ and $z_*$. The model with this function
$F$ gives rise to a dependence structure in the frequency-rank plot
described by $g(z) = \ln(z)$ before $z_*$ and $g(z)=\sqrt{z}$ after
$z_*$, and to a linear growth of $D_t$, that is a Heaps' law with
exponent $1$. (See Subsec.~\ref{sqrt-case} for technical details).

In Fig.~\ref{sim-double-lin} we exhibit the results for
\begin{equation}\label{F-double-lin}
F(x)=\begin{cases}
\rho_1  x  \quad&\mbox{for } 1\leq x<z_*\\
\rho_2x+ (\rho_1-\rho_2)z_*  \quad&\mbox{for } x\geq z_*\,,
\end{cases}
\end{equation}
for different values of $\rho_1,\,\rho_2$ and $z_*$. The model with
this function $F$ gives rise to a Zipf's law with two different
coefficients and to different kinds of behavior for $D_t$, depending
on the value $\rho_2/\nu$. (See Subsec.~\ref{double-lin-case} for technical
details).

Finally, in Fig.~\ref{sim-xlog} we exhibit the results for
\begin{equation}\label{F-xlog}
F(x)=\rho_1 x\ln(x+1) 
\end{equation}
for different values of $\rho_1$. The model with this update function
$F$ gives rise to an exponential decay of the frequency-rank function
$z(r)$ (that corresponds to $g(z)=\ln(\ln(z))$) and to a logarithm
growth of $D_t$. (See Subsec.~\ref{xlog-case} for technical
details). It is worthwhile to note that these behaviors for $z(r)$ and
$D_t$ can be achieved also by the standard urn with triggering with
$\nu=0$ (see\cite{Tria1}). However, the two models are completely
different: for the standard model, the assumption regards the
parameter $\nu$ that rules the probability $b_t$ of a new color, while
the number of balls of an old color increases linearly according to a
``free'' coefficient; whereas, for the proposed new model, the
assumption regards the update function $F$ that drives the growth of
the number of balls of an old color in the urn and the parameter $\nu$
can be whatever. \\

\begin{figure}[htbp]
\begin{center}
\includegraphics[width=0.47\textwidth]{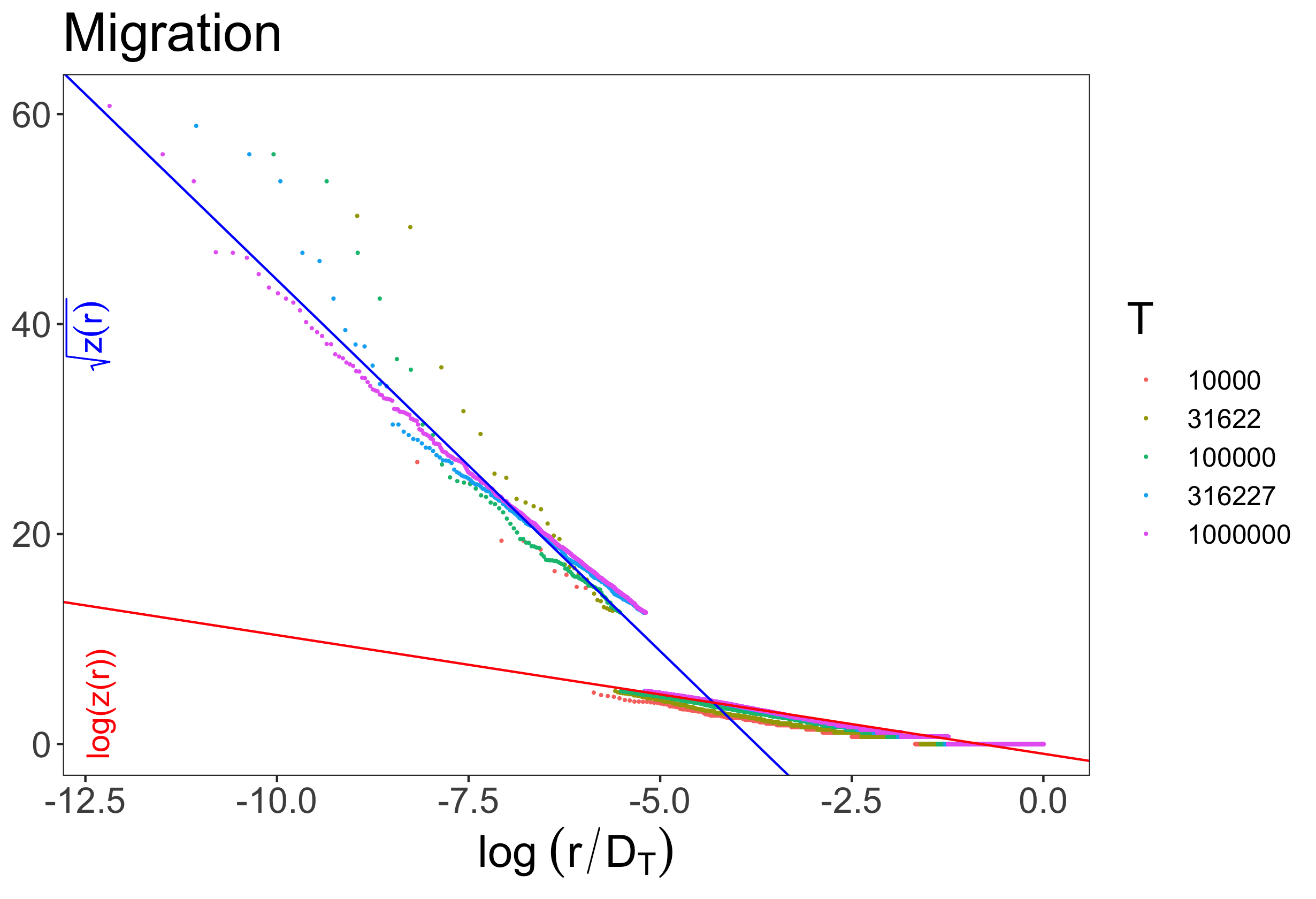}
 \includegraphics[width=0.47\textwidth]{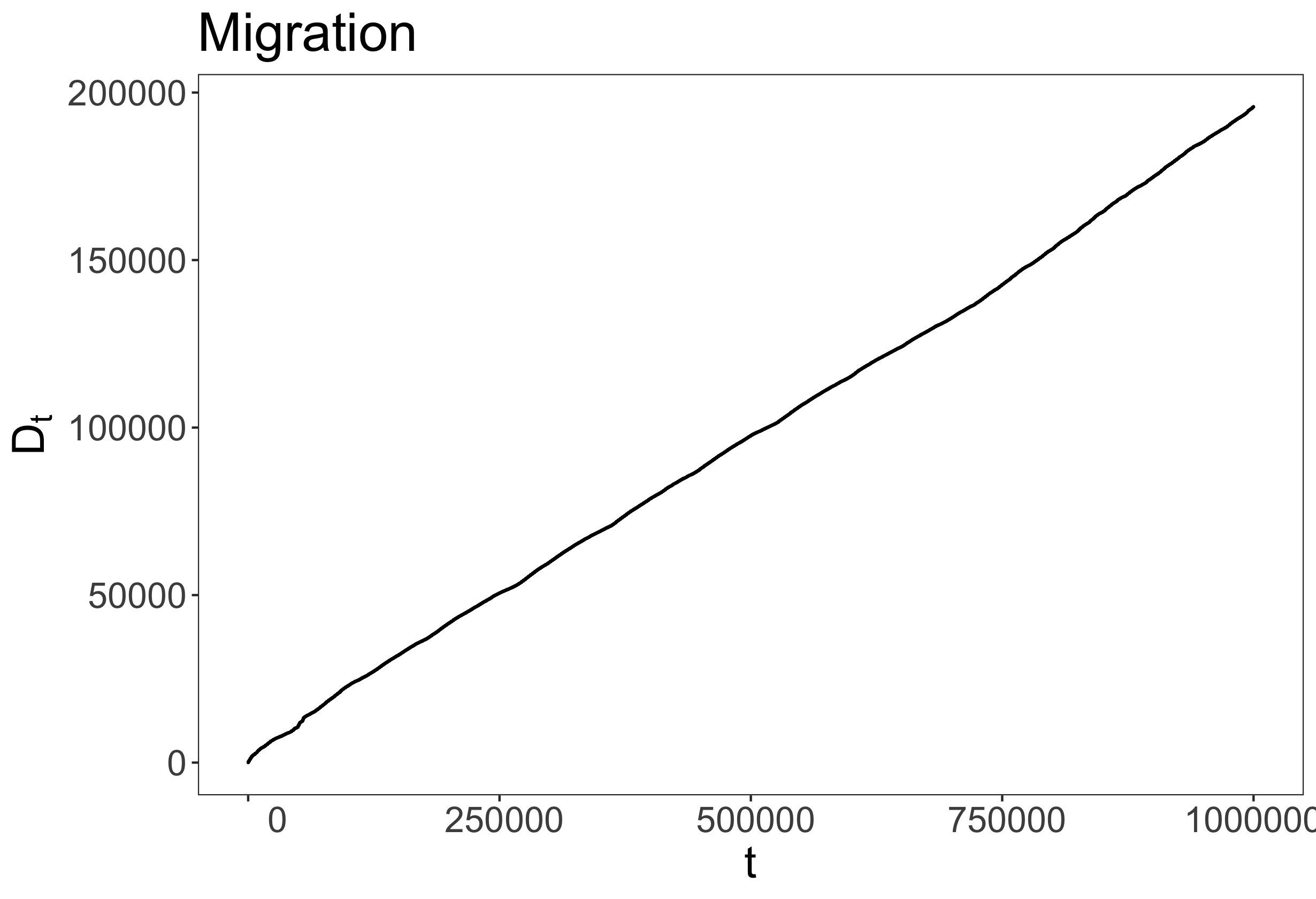}
\end{center}
	\caption{Migration: (Left) Frequency-rank plot in two
          different scales: log-sqrt before a certain $R_*$ and
          log-log after $R_*$. Parameters (see
          Subsec.s~\ref{sqrt-case} and \ref{app-estimation} for
          details): $a_1=7.07$, $a_2=1.13$, $R_*\in [-5.97,-5.19]$,
          that give $\rho/\nu=a_2=1.13$ and $z_*=4(a_1/a_2)^2=
          156.58$. The different colors of the dots correspond to
          different quantities of data taken for depicting the
          plot. (Right) Behavior of $D_t$.}
\label{migrants}
\end{figure}

\begin{figure}[htbp]
  \begin{center}
  \includegraphics[width=0.47\textwidth]{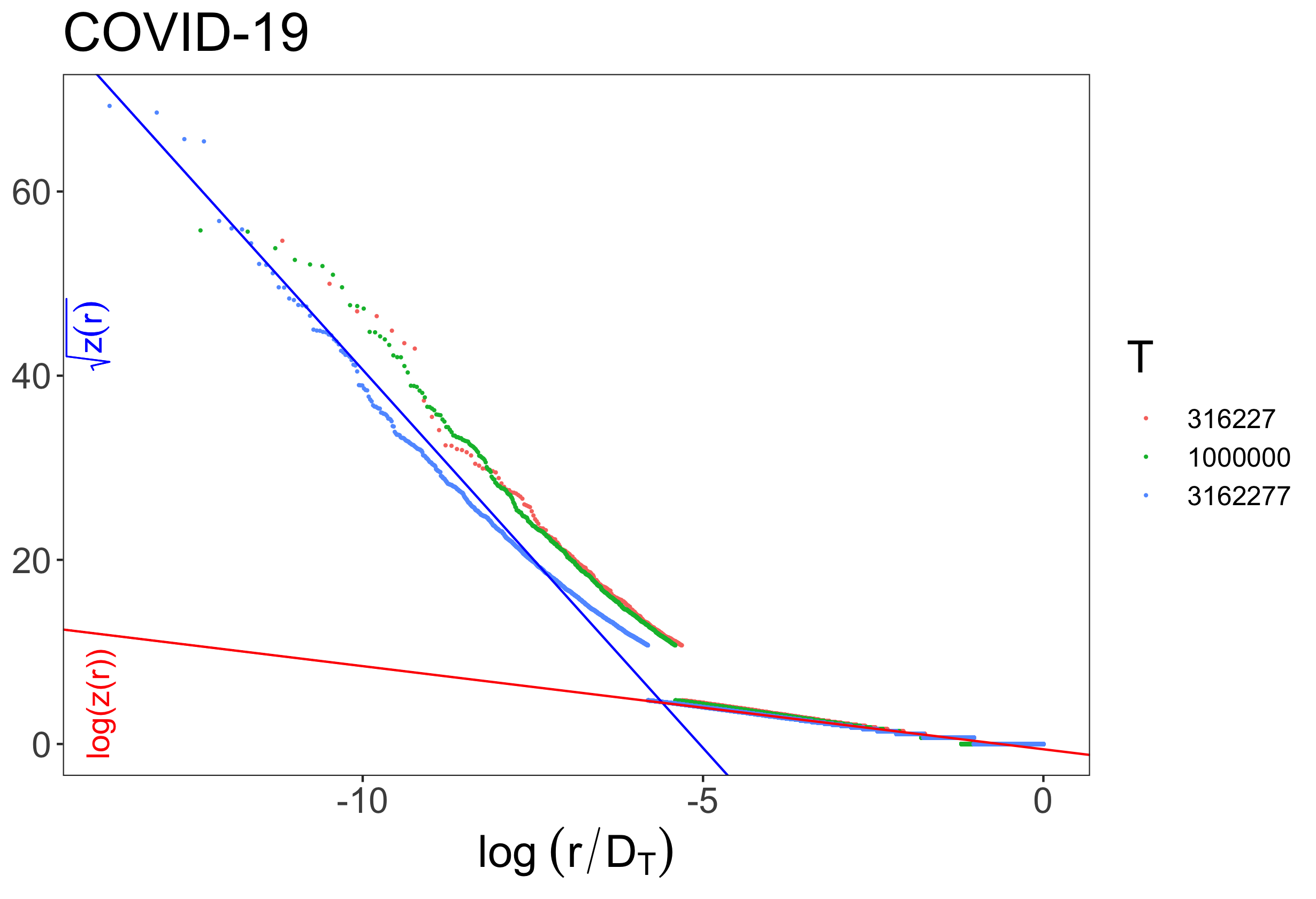}
  \includegraphics[width=0.47\textwidth]{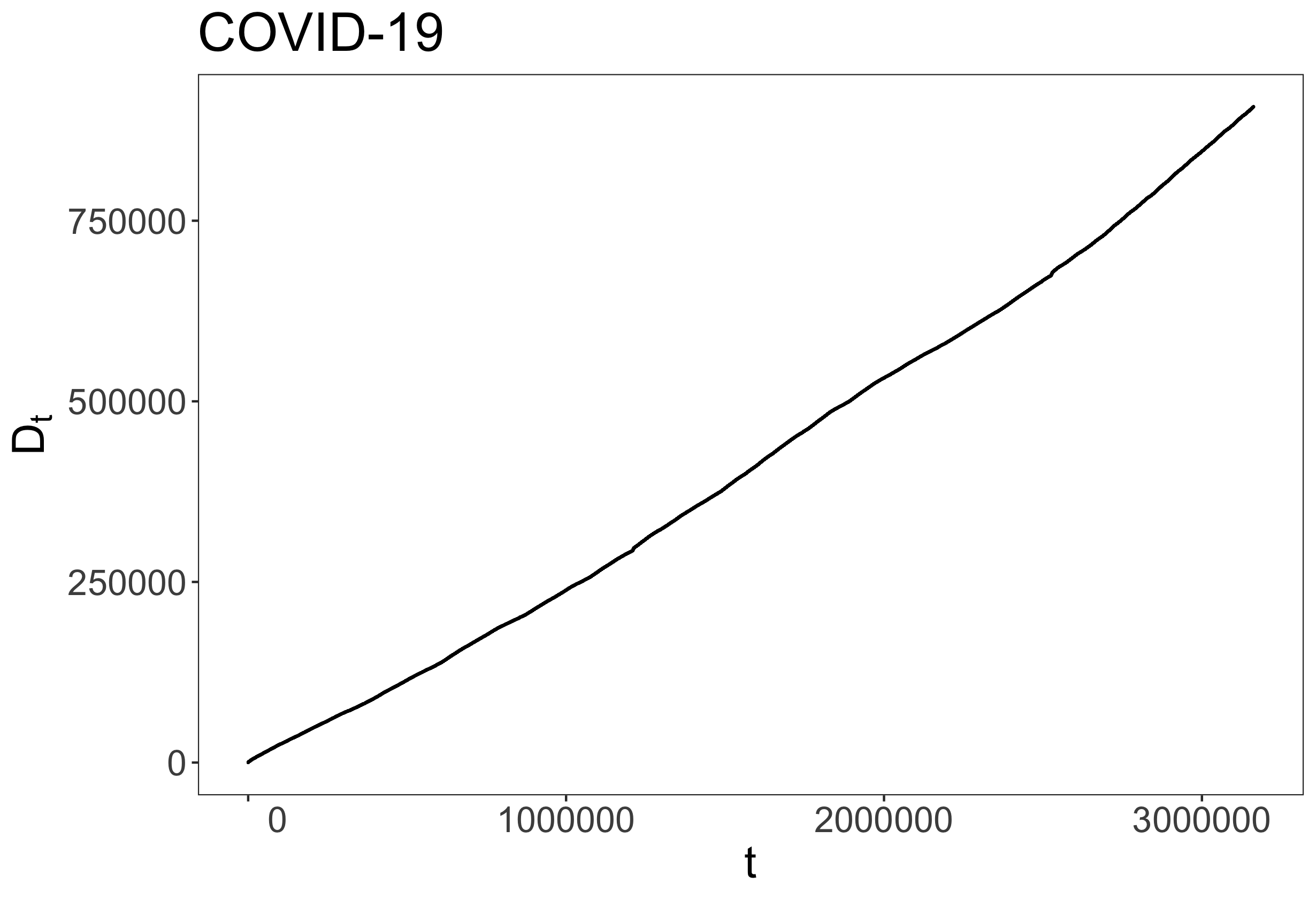}
\end{center}
  \caption{COVID-19: (Left) Frequency-rank plot with two different
    scales: log-sqrt before a certain $R_*$ and log-log after
    $R_*$. Parameters (see Subsec.s~\ref{sqrt-case} and
    \ref{app-estimation} for details): $a_1=8.68$, $a_2=0.895$,
    $R_*\in [-7.44,-6.80]$, that give $\rho/\nu=a_2=0.895$ and
    $z_*=4(a_1/a_2)^2= 376.23$. The different colors of the dots
    correspond to different quantities of data taken for depicting the
    plot. (Right) Behavior of $D_t$.}
\label{covid}
\end{figure}

\begin{figure}[htbp]
  \begin{center}
    \includegraphics[width=0.47\textwidth]{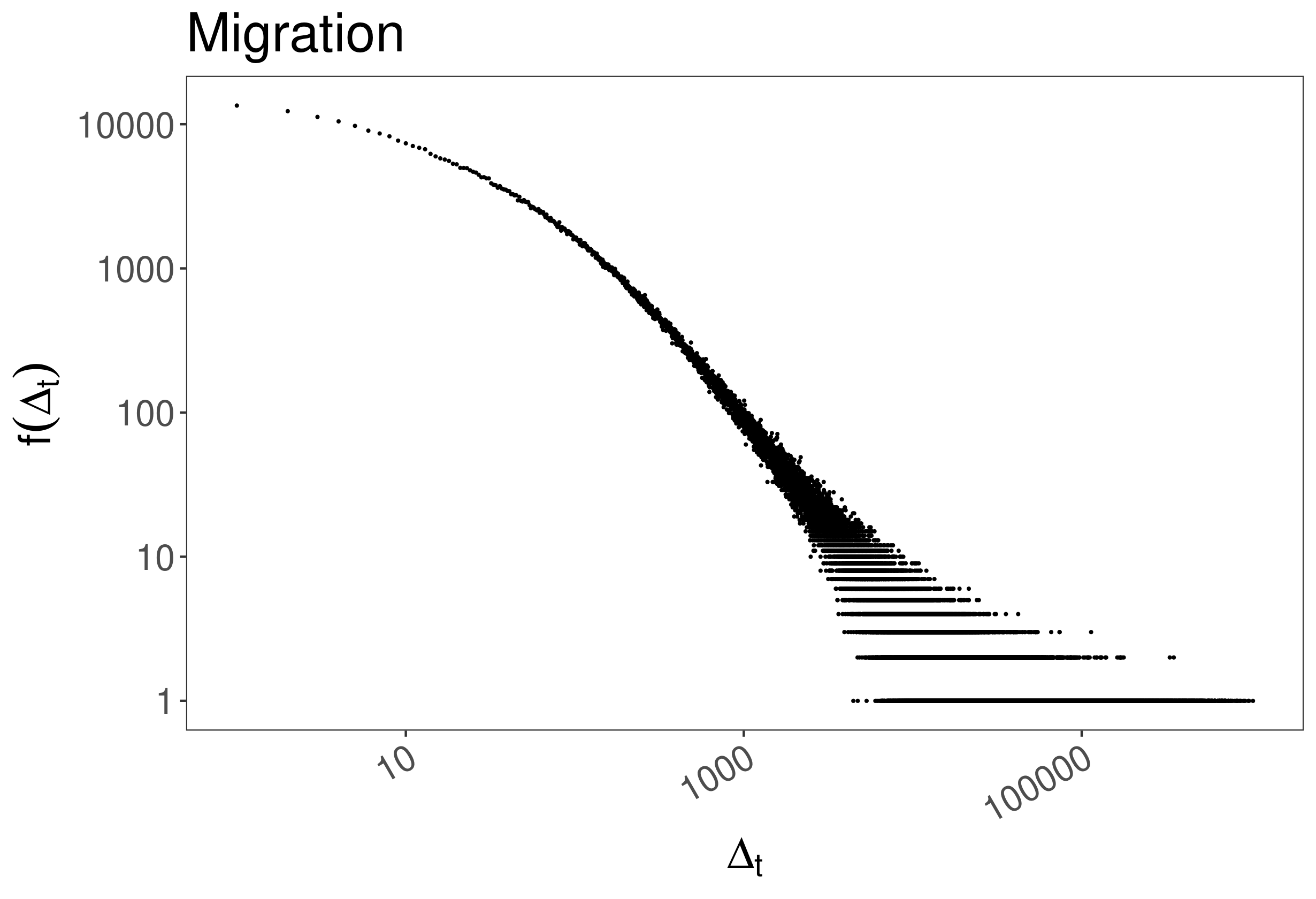}
    \includegraphics[width=0.47\textwidth]{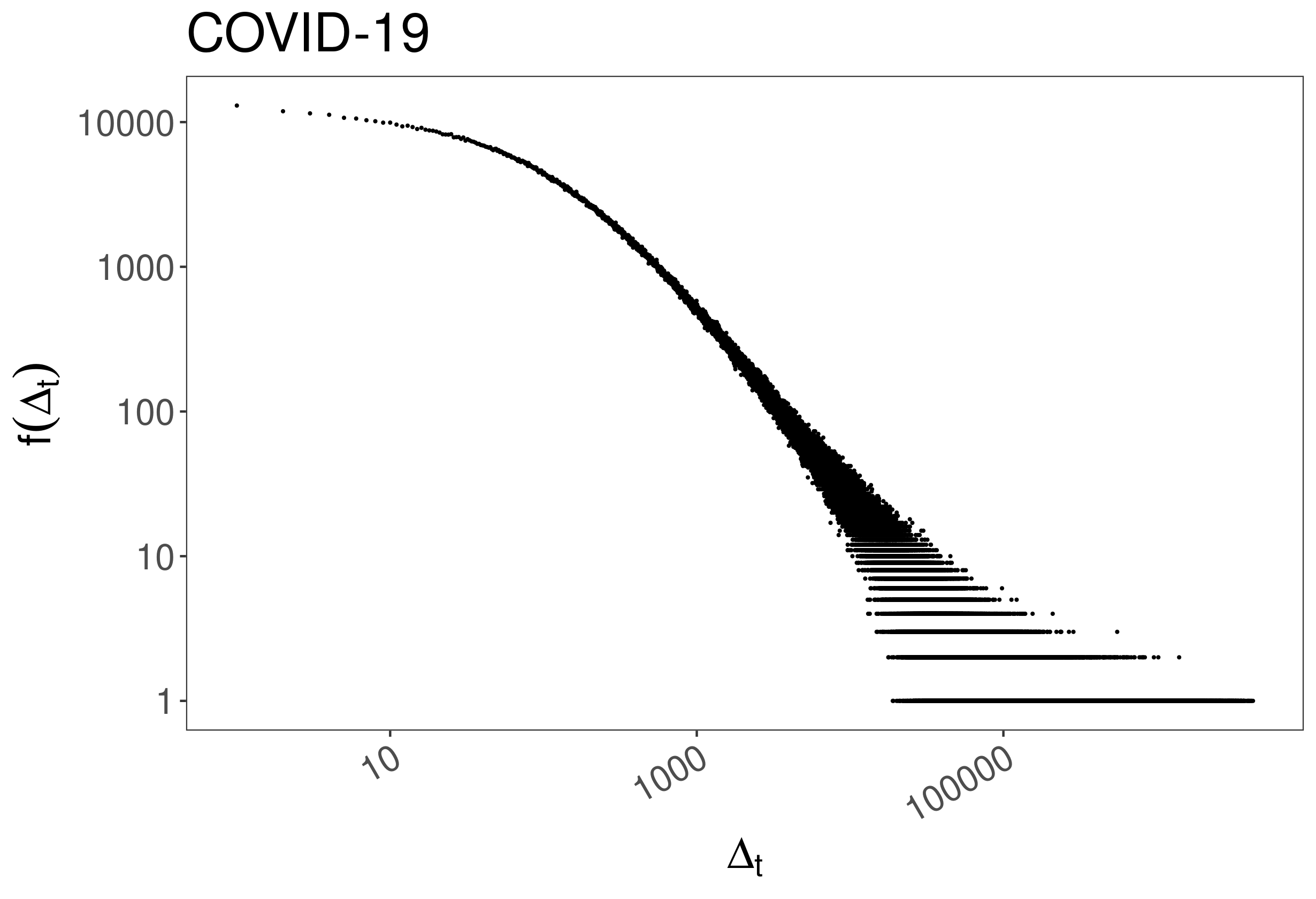}
    \\
    \includegraphics[width=0.47\textwidth]{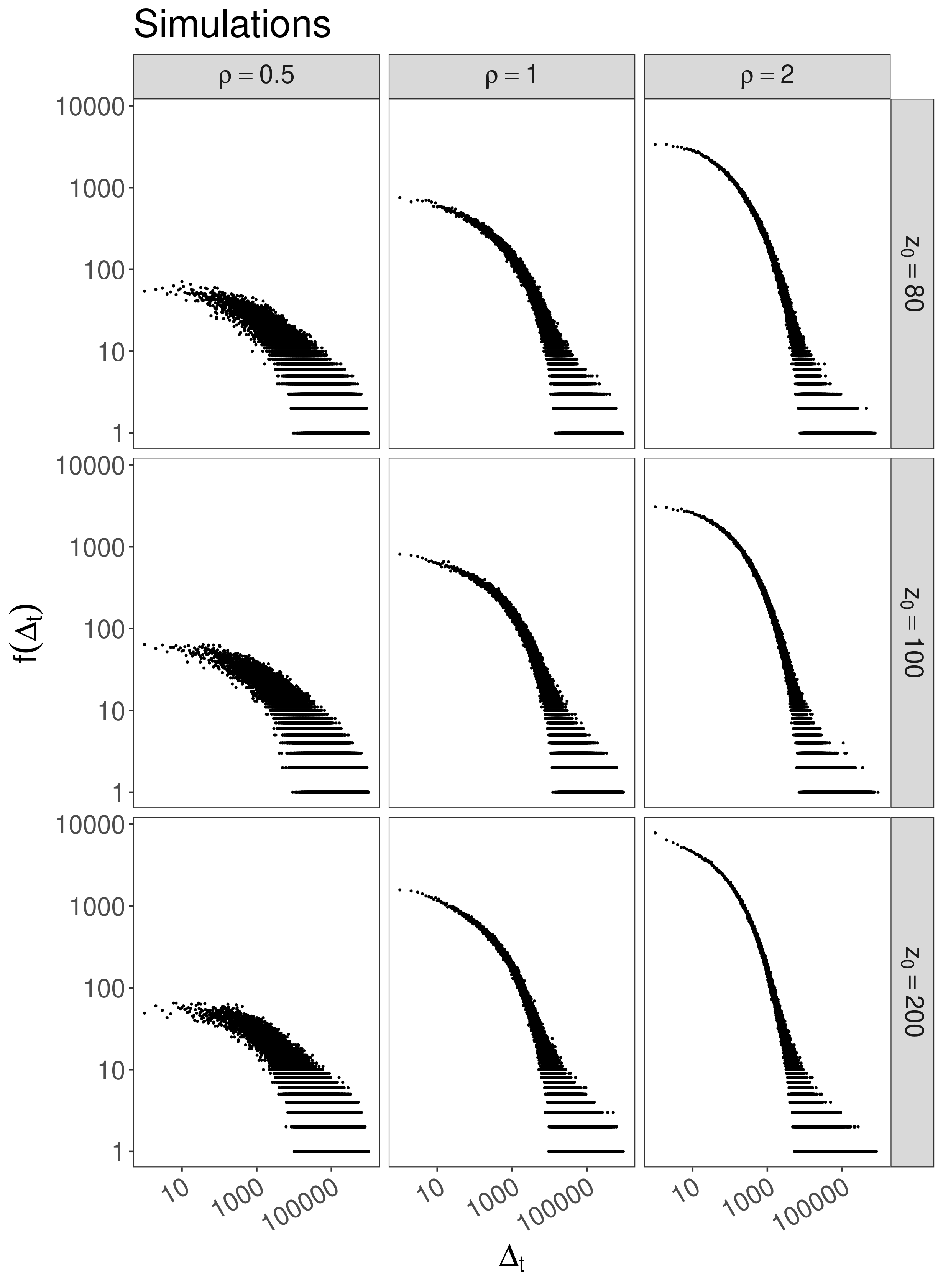}
\end{center}
  \caption{Frequency distribution of inter-event time steps between
    pairs of consecutive occurrences of the same color: (Up, Left)
    Migration; (Up, Right) COVID-19; (Below) Simulations of the model
    with $F$ given by \eqref{F-sqrt-gen}, $N_0=2$, $\nu=0.75$,
    $\widehat{\rho}=\rho$ and different values of $\rho$ and
    $z_*=z_0$.}
\label{interTimes}
\end{figure}

\noindent{\bf Empirical results.} Two data sets have been collected
from the Twitter platform, using the official API to stream the
exchange of messages on several topics:

\begin{itemize}
\item {\it Italy, Migration debate}\\ Data were collected through the
  Filter API since 23rd of January to 22nd of February 2019 and
  targeted the Italian debate on migration. The total number of posts
  is $T=1\,066\,677$. More details on the data set can be found
  in~\cite{Caldarelli2020a}.
%
%
\item{\it Italy, COVID-19 epidemic}\\ The data set covers the period
  from February 21st to April to 20th 2020, including tweets in
  Italian language. The keywords used for the query are relative to
  the COVID-19 epidemic. The total number of posts is
  $T=4\,580\,781$. More details on the data set can be found
  in~\cite{caldarelli2020analysis}.
%
%
\end{itemize}
Using the metaphor of the urn, the extractions correspond to the
publication of posts on Twitter. Therefore the sequence $\mathcal S$
of colors is constructed looking at the posts ordered by their
time-stamps. The color of the ball tells if the post is a new tweet
(that is a novelty) or a re-tweet/quote/reply (that is a repetition):
a new color is associated to a new tweet; while an old color
corresponds to a re-sharing (by a re-tweet or a quote or a reply) of
an old post. More precisely, in the latter case, we register in
$\mathcal{S}$ the color of the original message: for instance, given
$t_1<t_2<t_3$, if at time $t_3$ we have the quote of the post
published at time $t_2$, which is a retweet of the post at time $t_1$,
we register at positions $t_2$ and $t_3$ the same color of $t_1$. For
the sequence obtained from the data set regarding the migration
debate, the number of observed distinct colors is $D_T=210\,190$ and
the maximum number of times a given color is repeated is
$z_{max}=3\,694$. For the sequence obtained from the data set
regarding the COVID-19 epidemic, the number of observed distinct
colors is $D_T=1\,447\,623$ and the maximum number of times a given
color is repeated is $z_{max}=4\,818$. \\ \indent When a new tweet has
been sent, the addition of $\nu+1$ balls of new distinct colors can be
seen as the potential new tweets that the posted tweet may
generate. Hence the parameter $\nu$ is related to the ability of a
generic new tweet to give rise to future new tweets. On the other
hand, the update function $F$ rules the probability of a generic
posted tweet to be re-shared (with a retweet or a quote or a reply).
For all the considered data sets, we observe the same damping effect
on the old elements: the update function $F$ increases linearly until
a certain threshold, then it increases sub-linearly as the square
root. Indeed, looking at the empirical frequency-rank plot, we observe
a dependence structure given by $g(z) = \ln(z)$ before a certain
threshold $z_*$ and $g(z)=\sqrt{z}$ after $z_*$ (see the left panels
in Fig.s \ref{migrants} and \ref{covid}), that corresponds to the
update function $F$ in the model described in
\eqref{F-sqrt-gen}. Moreover, we verify (see the right panels in Fig.s
\ref{migrants} and \ref{covid}) the linear growth of the number $D_t$
of distinct observed tweets, which agrees with the proven theoretical
result (see Subsec.~\ref{sqrt-case}). Finally, Fig.~\ref{interTimes}
shows that for both data sets the frequency distribution $f(\Delta_t)$
of inter-event time steps $\Delta_t$ between pairs of consecutive
occurrences of the same color in $\mathcal S$ exhibits a behavior
similar to the one obtained by simulations of the model with $F$ given
by \eqref{F-sqrt-gen}.

%
%

%

%
%

%
  
\section{Discussion}\label{sec-discussion}
The innovation models introduced so far satisfy the Heaps' law,
regarding the rate at which novelties appear, and the Zipf's law, that
states a power law behavior for the frequency distribution of the
elements. However, there are empirical cases far from showing a pure
power law behavior and such a deviation is present for elements with
low ranks, that is with high frequencies. In this work we explain such
deviations from the Zipf's law by adding a suitable {\em damping effect} 
in the urn with triggering model. More precisely, we generalize the
standard urn with triggering \cite{Tria3, Tria1, Tria2} by the
introduction of a function $F$ that drives the {\em update mechanism}
of the number of balls of the same color of the extracted one when it
is of an old color. Indeed, if we take the update function $F$ linear
until a certain point and then still linear but with a smaller slope
or sub-linear, then we obtain a frequency-rank plot closer to the
empirical ones also in the part of high frequencies. This means that
the number of balls of an old color increases linearly with the number
of times it is extracted until a certain threshold, then it increases
slower, that is we have a damping factor in the updating of the urn.
\\

\indent Given the function $g$ that fits the empirical frequency-rank
plot (see eq.~\eqref{eq:Zipf}), we are able to find the corresponding
update function $F$ of the proposed generative model (see
eq.\eqref{rel-F-g}). This is a very useful result for applicative
purposes, since in applications one usually observes and tries to fit
the empirical frequency-rank plot. Further, we have shown how to
obtain the asymptotic behavior of the number $D_t$ of distinct
observed elements starting from the function $g$ and we have employed
this methodology with some specific functions $g$.  The obtained
theoretical results are supported by simulations.  \\

\indent We have applied the proposed model to some data sets from the
social platform Twitter, where the update function $F$ rules the
probability of a generic posted tweet to be re-shared. For all the
considered data sets, we observed the same damping effect on the old
elements: the update function $F$ grows linearly until a certain
threshold, then it increases sub-linearly, precisely according to the
square root. Moreover, we empirically verified the linear growth of
the variable $D_t$, which agrees with the proven theoretical result.
\\

\indent Finally, we underline that the proposed model provides a
general framework that is able to explain also the power law behavior
with two different scaling exponents observed in \cite{cancho, ger,
  petersen} and other kinds of empirically observed curves
(e.g. \cite{lu2013}). Therefore, it results a very flexible generative
model that could perfectly reproduce the frequency-rank plot for low
and high ranks, together with the behavior of $D_t$, in many contexts.

\section{Methods}Take $\nu>0$ and assume $F$ to be extended with continuity on the
whole $[1,+\infty)$ and in such a way that it is differentiable
  everywhere except in a finite number of points. 
 
\subsection{Relation between $F$ and the frequency distribution}
For each $k\geq 1$, we denote by $Q_{k,t}$ the number of colors $c$ in
$\mathcal S$ with $K_{c,t}=k$ and we set $p_k=\lim_t Q_{k,t}/D_t$. The
family $(p_k)_k$ is the (stationary) frequency distribution.  \\

\indent We have $ D'_t=b_t=\frac{N_0+\nu D_t}{T_t}$ and so we can
write $ T_t\approx \frac{\nu D_t}{D'_t}$. Moreover, we have
$\sum_{k\geq 1} Q_{k,t}=D_t$ and we can write the following master
equation for $Q_{k,t}$:
\begin{equation*}
  \begin{split}
  \frac{\partial Q_{k,t}}{\partial t}&=
  -\frac{Q_{k,t}F(k)}{T_t}
  +\frac{Q_{k-1,t}F(k-1)}{T_t}=
  -\frac{(Q_{k,t}-Q_{k-1,t})F(k)+Q_{k-1,t}(F(k)-F(k-1))}{T_t}
  \\
  &\approx - \frac{1}{T_t}\frac{\partial F(k)Q_{k,t}}{\partial k}
  \approx -\frac{D'_t}{\nu D_t}\frac{\partial F(k)Q_{k,t}}{\partial k}\,.
  \end{split}
\end{equation*}
 Using the asymptotic relations
$Q_{k,t}\approx p_k D_t$ and $\frac{\partial Q_{k,t}}{\partial
  t}\approx D'_t p_k$, from the above relation we get
  \begin{equation*}
p_k\approx -\frac{1}{\nu}\frac{d[F(k)p_k]}{dk}\,.
  \end{equation*}
Since  $\frac{d[F(k)p_k]}{dk}=F'(k)p_k+F(k)p'_k$, we obtain
$$
\frac{p'_k}{p_k}\approx -\frac{\nu}{F(k)}-\frac{F'(k)}{F(k)}\,.
$$ When $1/F$ has primitive function $H$, this equation has solution
$\ln(p_k)\approx -\ln(F(k))-\nu H(k) + C$, that is
\begin{equation}\label{eq:master2}
\ln(p_k) \approx - \ln(F(k))  - \nu\int_{k_0}^{k} \frac{1}{F(x)}dx +C'\,,
\end{equation}
where $C$ and $C'$ denote suitable constants. The above relation
provides the relationship between the function $F$ in the model and
the frequency distribution. For instance, for the standard urn with
triggering, we get $\ln(p_k)\approx
-(1+\nu/\rho)\ln\big(\widehat\rho+\rho(k-1)\big)+C''$, that is
$p_k\propto k^{-(1+\nu/\rho)}$ for large $k$ (see also \cite{Tria1}).


\subsection{Relation between $F$ and the frequency-rank plot}\label{app-g}
Assume a dependence in the frequency-rank plot of the form
\begin{equation}\label{eq:Zipf}
g(z(r)) = -a \ln(r) + b
\end{equation}
with a strictly increasing function $g$ and a constant $a>0$. Then, we
get
\begin{equation*}
r =  \exp \Big( -\frac{g(z(r))-b}{a} \Big)\,.
\end{equation*}
If $g$ is differentiable, we have 
\begin{equation*}
\frac{\partial z(r)}{\partial r} =
\frac{\partial g^{-1} (-a \ln(r) + b) }{\partial r} =
-\frac{1}{g'(g^{-1}(-a\ln(r)+b))} \frac{a}{r}\,.
\end{equation*}
Since $\delta r \approx p(z(r)) |\delta z|$, we get
\begin{equation*}
p(z(r)) \propto -\frac{1}{\frac{\partial z}{\partial r}}=
g'(g^{-1}(-a\ln(r)+b))\frac{r}{a}=
\frac{1}{a}g'(z(r))\exp \Big( -\frac{g(z(r))-b}{a} \Big).
\end{equation*}
Setting $y = z(r)$ in the above relation and taking the logarithm, we
find
\begin{equation}\label{eq:master3}
\begin{split}
\ln(p(y)) & \approx \ln(g'(y)) - \frac{g(y)}{a} + C_1
\\
& \approx - \ln\Big(\frac{a\nu}{g'(y)}\Big) - \nu \int_{y_0}^{y}
\frac{1}{a\nu} g'(s) \,ds + C_2\,.
\end{split}
\end{equation}
If we compare this last equation with \eqref{eq:master2}, we arrives
to the relation \eqref{rel-F-g} between the function $F$ of the model
and the function $g$ describing the frequency rank plot.


\subsection{Behavior of $D_t$}
Since \eqref{rel-Dt}, the behavior of $D_t$ can be obtained starting
from the frequency-rank function $r\mapsto z(r)$. For instance, in the
case of a pure (generalized) Zipf's law with $\alpha\neq 1$, we have
$$ t\approx \int_1^{D_t}
z_{max}r^{-\alpha}dr=z_{max}\frac{(D_t^{1-\alpha}-1)}{1-\alpha}
$$ and so, taking into account that $z(D_t)\propto 1$,
i.e. $z_{max}\propto D_t^{\alpha}$, we get $D_t\propto t^{1/\alpha}$
for $\alpha<1$ and $D_t\propto t$ for $\alpha>1$. When $\alpha=1$, we
find $t\approx D_t \ln(D_t)$ and so $D_t\propto t/\ln(t)$.
\\

\indent When $z(r)$ exhibits two different behaviors, one for small
ranks, say for $r< \xi_t$, and the other for large ranks, say for $r>
\xi_t$, the above relation becomes
\begin{equation}\label{rel-for-Dt}
t\approx \int_1^{\xi_t}z(r)dr +\int_{\xi_t}^{D_t}z(r)dr\,.
\end{equation}
In the following, we will study different cases: the first one is
observed in the real data sets that we discuss in the present work and
the other two have been observed in other papers \cite{cancho, ger,
  lu2013, petersen}.
%
%

\subsubsection{$g(z)=\ln z$ for $z<z_*$ and $g=\sqrt{z}$ for $z>z_*$}
\label{sqrt-case}
Suppose that the frequency-rank plot identifies the following
dependence structure:
\begin{equation}\label{eq:zr_double_a}
\begin{cases}
\sqrt{z(r)} - \sqrt{z_*} =
 -a_1 \big[  
 \ln\big( \frac{r}{D_t} \big) -R_* 
  \big]
  &\mbox{for }  
  \ln \big( \frac{r}{D_t} \big) < R_*
  \\
\ln{z(r)} - \ln{z_*} =
  -a_2 \big[  
  \ln\big( \frac{r}{D_t} \big) -R_* 
  \big]
  &\mbox{for } 
  \ln \big( \frac{r}{D_t} \big) > R_*\,,
\end{cases}
\end{equation}
where $a_1,\,a_2,\, R_*$ and $z_*$ are constants such that $a_i>0$,
$R_*<0$ and $z_*=z(\xi_t)$ with $\xi_t=D_te^{R_*}$. This corresponds
to
\begin{equation*}
g(z) = 
\begin{cases}
\ln(z) &\mbox{for } z < z_*\\
\sqrt{z}  &\mbox{for } z > z_*
\end{cases}
\end{equation*}
and, leveraging \eqref{rel-F-g}, we detect the function $F$ in the model as
\begin{equation}\label{F-sqrt-gen}
F(x)=\begin{cases}
\widehat{\rho}-\rho + \rho  x \quad&\mbox{for } 1\leq x<z_*\\
\widehat{\rho}-\rho+ \rho \sqrt{z_*x}  \quad&\mbox{for } x\geq z_*\,,
\end{cases}
\end{equation}
where $\widehat{\rho}>0$, $\sqrt{z_*}\rho = 2\nu a_1$, $\rho=\nu
a_2>0$ and so
\begin{equation}\label{eq:estimation_z*}
z_*=4\left(\frac{a_1}{a_2}\right)^2.
\end{equation}
\indent From \eqref{eq:Zipf} and
\eqref{eq:zr_double_a}, we get
\begin{equation*}
\begin{aligned}
  z(r) & =\begin{cases}
g^{-1}(-a_1 \ln r + b_1) &\mbox{for }  \ln \big( \frac{r}{D_t} \big) < R_*\\
g^{-1}(-a_2 \ln r + b_2) &\mbox{for }  \ln \big( \frac{r}{D_t} \big) > R_*
\end{cases}
\\
& =
\begin{cases}
(-a_1 \ln r + b_1)^2 &\mbox{for } r < \xi_t = D_t e^{R_*} 
\\
e^{b_2} r^{-a_2}   &\mbox{for } r > \xi_t = D_t e^{R_*} \,,
\end{cases}
\end{aligned}
\end{equation*}
where
%
%
\begin{equation}\label{eq:zr_double_c}
  -a_1 \ln \xi_t + b_1 = \sqrt{z_*}\qquad\mbox{and}\qquad
  -a_2 \ln \xi_t + b_2 = \ln(z_*).
\end{equation}
Now, we recall that 
$$
\int_1^x (-a_1\ln(r)+b_1)^2dr=
x\left[(-a_1\ln(x)+b_1)^2 + 2a_1(-a_1\ln(x)+a_1+b_1)\right]
-b_1^2-2a_1^2-2a_1 b_1\,.
$$
Taking in the above integral $x=\xi_t$ and using the first equality
in \eqref{eq:zr_double_c},  we obtain
\begin{equation}\label{eq:first-int}
\int_1^{\xi_t}z(r)dr= \xi_t  (z_*+2a_1^2+2a_1\sqrt{z_*})
+ O(\ln^2(\xi_t)) = D_t e^{R_*}  z_*[1+\rho^2/(2\nu^2)+\rho/\nu] 
+ O(\ln^2(D_t)) \,.
\end{equation}
For the second integral, assuming $a_2=\rho/\nu\neq 1$ and using the
second equality in \eqref{eq:zr_double_c}, we get 
\begin{equation}\label{eq:second-int}
  \begin{split}
    \int_{\xi_t}^{D_t}z(r)dr&=
    \frac{e^{b_2}}{1-a_2} [ D_t^{1-a_2} - \xi_t^{1-a_2} ]
    = D_t \frac{z_* }{1-a_2}(  e^{a_2R_*} - e^{R_*}) 
    \\
    &=  D_t \frac{z_* e^{R_*}}{(\rho/\nu-1)} ( 1 - e^{R_*(\rho/\nu-1)})  
\,,
  \end{split}
\end{equation}
while for $a_2 = \rho/\nu=1$, we get 
\begin{equation}\label{eq:second-int2}
    \int_{\xi_t}^{D_t}z(r)dr=
    e^{b_2} (\ln(D_t) - \ln(\xi_t) )= -z_* R_* D_t e^{R_*}\,.
\end{equation}
From \eqref{rel-for-Dt}, \eqref{eq:first-int}, \eqref{eq:second-int}
and \eqref{eq:second-int2}, we can conclude that $D_t\propto t$. This
means that we have the Heaps' law with exponent $\gamma=1$.


\subsubsection{Zipf's law with two different coefficients}
\label{double-lin-case}
Suppose that the frequency-rank plot identifies a ``double'' Zipf's
law, that is the following dependence structure:
\begin{equation}\label{eq:zr_double_second}
\begin{cases}
\ln{z(r)} - \ln{z_*} =
 -a_1 \big[  
 \ln\big( \frac{r}{D_t} \big) -R_* 
  \big]
  &\mbox{for }  
  \ln \big( \frac{r}{D_t} \big) < R_*
  \\
\ln{z(r)} - \ln{z_*} =
  -a_2 \big[  
  \ln\big( \frac{r}{D_t} \big) -R_* 
  \big]
  &\mbox{for } 
  \ln \big( \frac{r}{D_t} \big) > R_*\,,
\end{cases}
\end{equation}
where $a_1,\,a_2$ and $R_*$ are constants such that $a_i>0$, $a_1\neq
a_2$ (typically $a_1<a_2$), $R_*<0$ and $z_*=z(\xi_t)$ with
$\xi_t=D_te^{R_*}$. This kind of dependence was observed
in~\cite{cancho, ger, petersen} and, according to our model (see
\eqref{rel-F-g}), it corresponds to
$$
F(x)=\begin{cases} 
\widehat{\rho}-\rho_1+\rho_1 x  \quad&\mbox{for } 1\leq x<z_*\\
\widehat{\rho}-\rho_1+\rho_2 x + (\rho_1-\rho_2)z_*
\quad&\mbox{for } x\geq z_*\,,
\end{cases}
$$ where $\rho_1=\nu a_2\neq \rho_2 = \nu a_1$ (typically
$\rho_2<\rho_1$) and $\widehat{\rho}>0$.  \\ \indent From
\eqref{eq:Zipf} and \eqref{eq:zr_double_second}, we get
\begin{equation*}
\begin{aligned}
  z(r) & =\begin{cases}
g^{-1}(-a_1 \ln r + b_1) &\mbox{for }  \ln \big( \frac{r}{D_t} \big) < R_*\\
g^{-1}(-a_2 \ln r + b_2) &\mbox{for }  \ln \big( \frac{r}{D_t} \big) > R_*
\end{cases}
\\
& =
\begin{cases}
e^{b_1} r^{-a_1} &\mbox{for } r < \xi_t = D_t e^{R_*} 
\\
e^{b_2} r^{-a_2}   &\mbox{for } r > \xi_t = D_t e^{R_*} \,,
\end{cases}
\end{aligned}
\end{equation*}
where $b_1 = \ln{z_*} + a_1 \ln\xi_t $ and $b_2 = \ln{z_*} + a_2
\ln\xi_t $.  Therefore, we have
\begin{equation}\label{eq:first-int-second}
  \int_1^{\xi_t}z(r)dr= \begin{cases} \frac{z_*}{1-a_1}(D_t e^{R_*}-
    D_t^{a_1} e^{R_*a_1}) \qquad&\mbox{for } a_1=\rho_2/\nu\neq 1\\
    z_*e^{R_*}D_t(\ln(D_t)+R_*)\qquad&\mbox{for } a_1=\rho_2/\nu=1\,.
    \end{cases}
\end{equation}
  For the second integral, as before, we have 
  \begin{equation}\label{eq:second-int-second}
    \int_{\xi_t}^{D_t}z(r)dr= \begin{cases}
D_t\frac{z_* e^{R_*}}{a_2-1}(1-e^{R_*(a_2-1)})
\qquad&\mbox{for } a_2=\rho_1/\nu\neq 1\\
-z_*R_*D_t e^{R_*}\qquad&\mbox{for } a_2=\rho_1/\nu= 1\,.
    \end{cases}
  \end{equation}
  From \eqref{rel-for-Dt}, \eqref{eq:first-int-second},
  \eqref{eq:second-int-second}, we can conclude that the asymptotic
  behavior of $D_t$ is ruled by the value of $a_1=\rho_2/\nu$:
  $D_t\propto t$ when $a_1<1$, $D_t\propto t/\ln(t)$ when $a_1=1$
  and $D_t\propto t^{1/a_1}$ when $a_1>1$.


  \subsubsection{Frequency-rank plot with exponential decay}
  \label{xlog-case}
Suppose that the frequency-rank plot identifies the following
dependence structure:
\begin{equation}\label{eq:zr_exp}
\ln(z(r)) = -a r + b,\quad\mbox{that is}\quad z(r)=C e^{-ar}\,, 
\end{equation}
where $a>0$ and $C=e^b\propto e^{aD_t}$. This is the case observed in
\cite{lu2013}. The corresponding behavior of $D_t$ is given by the
relation \eqref{rel-for-Dt}, that is
$$
t\approx -\frac{C}{a}(e^{-a D_t}-e^{-a})=\frac{1}{a}(e^{a(D_t-1)}-1)\approx
\frac{1}{a}e^{aD_t}\,,
$$ that implies $D_t\propto \ln(t)$. These behaviors for $z(r)$ and
$D_t$ can be achieved by the introduced model with
$F(x)=\rho(x+1)\ln(x+1)$ and $\rho>0$. Indeed, starting from
\eqref{eq:zr_exp} and employing an adaption of the argument used in
Subsection \ref{app-g}, we find $p(z(r))\propto \frac{1}{a z(r)}$ and
so $\ln(p(y))\approx -\ln(y)$. On the other hand, when inserting the
above function $F$ in \eqref{eq:master2}, we find $\ln(p(k))\approx
-\ln(k+1)-(1+\nu/\rho)\ln(\ln(k+1))+C''\propto -\ln(k)$. The relation
\eqref{rel-F-g} is satisfied with $g(z)=\ln(\ln(z))$ and
$\rho=a\nu$.


\subsection{Empirical analysis: parameters estimation}
\label{app-estimation}

\begin{figure}[htbp]
  \begin{center}
\includegraphics[width=0.47\textwidth]{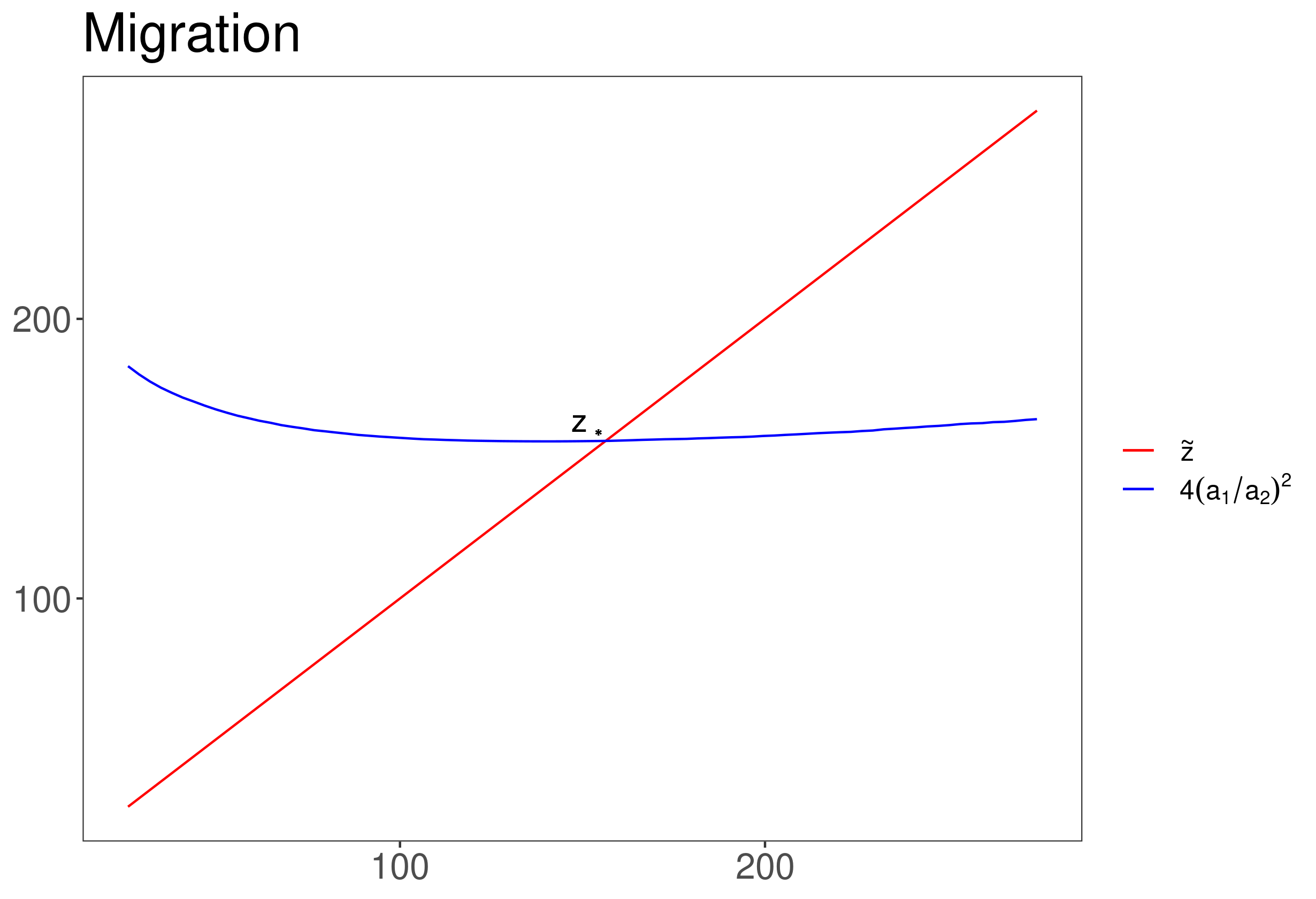}
\includegraphics[width=0.47\textwidth]{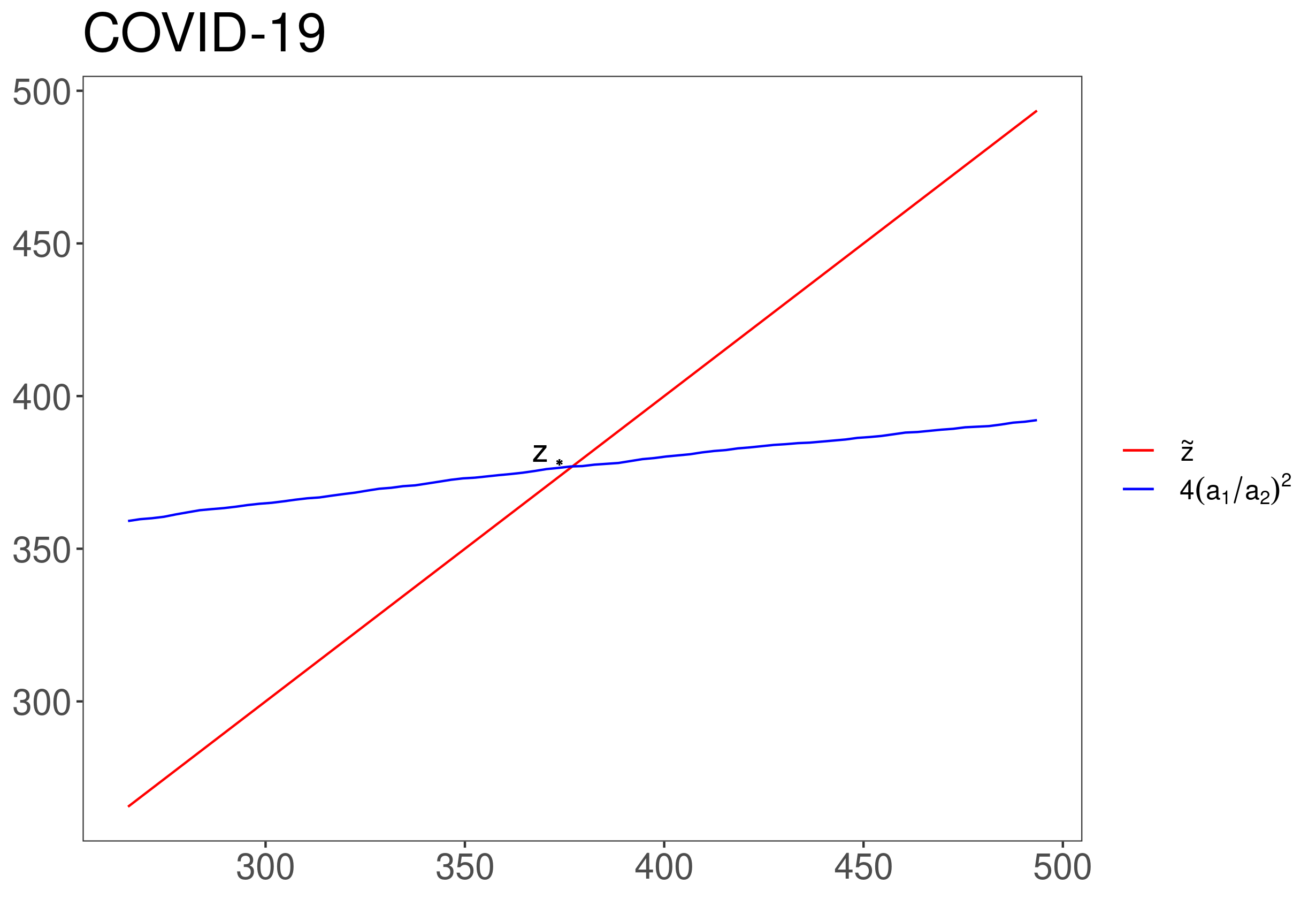}
\end{center}
  \caption{Estimation of the parameter $z_*$ based on
    Eq.~\eqref{eq:estimation_z*}: (Left) Migration; (Right) COVID-19.}
\label{estimation_z*}
\end{figure}

For each $\tilde{z}$, we fit 
\begin{equation*}
\begin{cases}
\sqrt{z(r)}=
 -a_1 \big[  
 \ln\big( \frac{r}{D_T} \big) + b_1 
  \big]
  &\mbox{for }  z(r)> \tilde{z}
  \\
\ln{z(r)} =
  -a_2 \big[  
  \ln\big( \frac{r}{D_T} \big) + b_2  
  \big]
  &\mbox{for } z(r)< \tilde{z}
\end{cases}
\end{equation*}
and we compute the corresponding quantity $4(a_1/a_2)^2$ (see
\eqref{eq:estimation_z*}). Finally, as shown in
Figure~\ref{estimation_z*}, we choose $z_*=\tilde{z}$ such that
$\tilde{z}=4(a_1/a_2)^2$ and we set $\rho/\nu$ equal to the
corresponding $a_2$.

\section*{Acknowledgements}
Both authors sincerely thank Fabio Saracco for having collected and shared with them the two Twitter data sets. 
\indent Giacomo Aletti is a member of the Italian Group ``Gruppo
Nazionale per il Calcolo Scientifico'' of the Italian Institute
``Istituto Nazionale di Alta Matematica'' and Irene Crimaldi is a
member of the Italian Group ``Gruppo Nazionale per l'Analisi
Matematica, la Probabilit\`a e le loro Applicazioni'' of the Italian
Institute ``Istituto Nazionale di Alta Matematica''.  \\[5pt]
\noindent{\bf Funding Sources}\\
\noindent Irene Crimaldi is partially supported by the Italian
``Programma di Attivit\`a Integrata'' (PAI), project ``TOol for
Fighting FakEs'' (TOFFE) funded by IMT School for Advanced Studies
Lucca.

\section*{Author contributions statement}

Both authors equally contributed to this work.

\section*{Additional information}

The author(s) declare no competing interests.

\end{document}